\documentclass[runningheads]{llncs}
\usepackage[T1]{fontenc}
\usepackage{graphicx}
\usepackage[table,xcdraw]{xcolor} 
\usepackage{pifont} 
\newcommand{\cmark}{\ding{51}}
\newcommand{\xmark}{\ding{55}}
\newcommand{\omark}{\ding{72}}
\usepackage{multirow}
\usepackage{float}
\floatstyle{plaintop}
\restylefloat{table}
\usepackage{dblfloatfix}
\newcommand{\tabitem}{~~\llap{\textbullet}~~}

\begin{document}
\title{Security Evaluation in Software-Defined Networks}
\author{Igor Ivki\'c\inst{1,2}\orcidID{0000-0003-3037-7813} \and
	Dominik Thiede\inst{2}\orcidID{0009-0001-2402-8740} \and
	Nicholas Race\inst{1}\orcidID{0000-0002-6870-8078} \and
	Matthew Broadbent\inst{3}\orcidID{0000-0002-7029-6893} \and
	Antonios Gouglidis\inst{1}\orcidID{0000-0002-4702-3942}
}
\authorrunning{I. Ivki\'c et al.}

\institute{Lancaster University, Lancaster, UK \and
	 University of Applied Sciences Burgenland, Eisenstadt, AT \and
	Edinburgh Napier University, Edinburgh, UK}
\maketitle              
\vspace{-15pt}
\begin{abstract}
Cloud computing has grown in importance in recent years which has led to a significant increase in Data Centre (DC) network requirements. A major driver of this change is virtualisation, which allows computing resources to be deployed on a large scale. However, traditional DCs, with their network topology and proliferation of network endpoints, are struggling to meet the flexible, centrally managed requirements of cloud computing applications. Software-Defined Networks (SDN) promise to offer a solution to these growing networking requirements by separating control functions from data routing. This shift adds more flexibility to networks but also introduces new security issues. This article presents a framework for evaluating security of SDN architectures. In addition, through an experimental study, we demonstrate how this framework can identify the threats and vulnerabilities, calculate their risks and severity, and provide the necessary measures to mitigate them. The proposed framework helps administrators to evaluate SDN security, address identified threats and meet network security requirements.
	
	\keywords{Software-Defined Networks \and Security Evaluation Framework \and Threat Analysis \and Risk Assessment \and Attack Modelling \and Threat Mitigation.}
\end{abstract}
\section{Introduction}
\label{sec:introduction}

Traditional communication networks and Data Centres (DCs) have relied on specialised, costly equipment, increasing the complexity and expense of establishing large-scale networks. Further, traditional network components (e.g., switches, routers, etc.) followed a "closed system" architecture, combining data and control planes. This system led to a restrictive management environment where only vendors could modify network configurations to meet customer requirements \cite{khondoker-2018}.

By separating the control plane from the data plane, Software-Defined Networks (SDNs) provide greater flexibility than traditional Data Centre Networks (DCNs). This separation shifts network control from the hardware devices to a logically centralised, software-based component called an SDN controller. The SDN controller, residing in the control plane, calculates centralised forwarding tables and pushes them to the network elements in the data plane.

\begin{figure*}[!h] \centering \includegraphics[width=4cm]{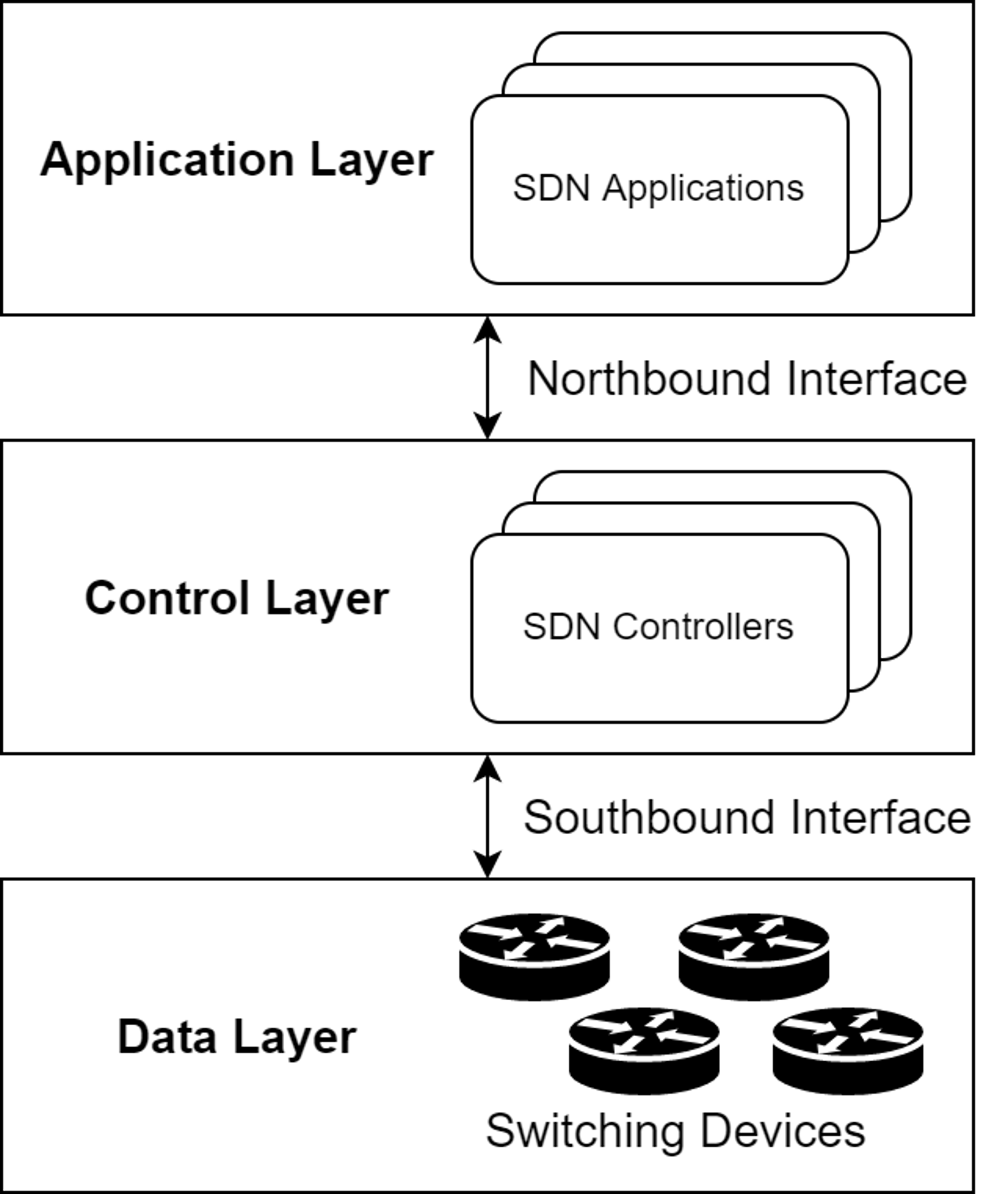} \caption{A Three-Layered SDN Topology consisting of a Data Layer, a Control Layer and an Application Layer (adapted from Xia et al., 2015).} \label{fig:fig1-1} \end{figure*}

The three-layer architecture shown in Figure \ref{fig:fig1-1} allows for a programmable network controller that provides an abstracted layer at which application and network services can be applied. This enables the network to be agile, cost-effective, manageable, and scalable to meet the growing network demands of demanding cloud applications \cite{correa-chica-2020}. However, there are also new security challenges that come with the new SDN architecture. Network intelligence is now centralised in a single (virtualised) SDN controller, compared to traditional network architectures with many distributed forwarding devices. This centralisation of network control is one of the most important benefits of SDN, but it is also one of its most significant weaknesses. While SDN enables administrators to manage the network from a central location, simplifying a holistic view of the network and faster decision making, a successful cyber-attack that compromises the SDN controller or application could affect the entire network \cite{al-saghier-2019}. Network security is a top priority in DC environments, where attacks especially on critical infrastructure could be catastrophic \cite{ivkic-2023}.

\vspace{-15pt} \label{sec:sdn-topology} 
\vspace{-15pt} 

In this article we present a Security Evaluation Framework for SDN components in DC environments. The framework comprises four phases, with each phase’s output used as input for the next. This article is an extended version from our previous work in \cite{ivkic-2023}, where we first introduced the framework. Compared to the previous publication \cite{ivkic-2023} where the focus was to introduce the framework and its four phases, this article provides more detailed information about the outputs of each phase. Furthermore, we present three central solutions and a detailed description of how they are implemented including a mapping of the threats they mitigate. The presented framework consists of four stages, where a threat and vulnerability analysis based on two different methodologies is performed first. These include the Spoofing, Tampering, Repudiation, Information Disclosure, Denial of Service (DoS), Elevation of Privilege (STRIDE) model and the Process for Attack Simulation and Threat Analysis (PASTA) approach \cite{pasta-2-2017}. Next, a risk and impact analysis is conducted including a ranking for each risk according to its severity using the Common Vulnerability Scoring System (CVSS) v3.1. The next stage uses the output from the previous analysis to model attack scenarios for the three highest ranked risks. The final stage verifies the calculated severity impact of each attack scenario in an experimental study using an SDN testbed based on Mininet \cite{kaur-mininet-2014} and an Open Network Operating System (ONOS) controller. We conclude with a presentation of countermeasures to either mitigate single threats by enabling integrated features or multiple threats by implementing a centralized solution \cite{ivkic-2023}.

The rest of this article is organised as follows: In Section \ref{sec:related_work} we discuss related work. We then introduce the Security Evaluation Framework and briefly describe its four stages in Section \ref{sec:framework}. We also present our experimental testbed, used in the subsequent sections (Section \ref{sec:Stage1} – \ref{sec:Stage4}), to conduct an experimental study. Each of these sections (Section \ref{sec:Stage1} – \ref{sec:Stage4}) represents one stage of the framework where we demonstrate how it works and discuss the output (results) of each stage in detail. In Section \ref{sec:Stage4}, we present three central solutions for mitigating multiple SDN threats. This section's output represents a correlation map of SDN threats, vulnerabilities, and mitigations. Finally, in Section \ref{sec:conclusion} we conclude our work.
\vspace{-15pt}
\section{Related Work}
\label{sec:related_work}
\vspace{-10pt}

In this section, we highlight the differences between existing studies and our work, detailing how the proposed framework addresses the gaps in the related research. Papers identified from the literature review are categorized into three areas: security analysis, security remediation, and new solution proposal. They are marked according to whether a Proof-of-Concept (PoC) test setup was utilized for validation and which SDN layer or interface was evaluated. Lastly, we discuss the difference between the SDN Security Evaluation Framework presented in this article and the related work.

\vspace{-15pt}
\subsection{Literature Review on Evaluating Security in SDN}

As the importance of SDN continues to grow in various research areas, an increasing number of security evaluations and tests have been carried out on this new architecture. For instance, Ruffy et al. used the STRIDE model to identify several SDN vulnerabilities and proposed mitigation solutions for each of the six STRIDE categories. They outlined a secure SDN design example for two network domains based on their findings. This design integrated multiple SDN controllers for redundancy and incorporated both traditional and new protection solutions. These solutions included Intrusion Detection Systems (IDS), firewalls and access control via Authentication, Authorisation and Accounting (AAA) servers. However, they may have overlooked security vulnerabilities that do not fall into the STRIDE categories, as their threat analysis considered only one approach based on six generic Threat Categories (TCs) \cite{ruffy-2016}.

Iqbal et al. highlighted several vulnerabilities and possible attacks in SDN, emphasizing their impacts and potential solutions. Their approach mainly focused on the communication channels within SDN layers. The proposed resolutions include Secure Socket Layer (SSL) / Transport Layer Security (TLS) data encryption, Advanced Encryption Standard (AES) and Data Encryption Standard (DES) ciphers, and role-based authorization derived from the FortNOX3 instance. They also provided a correlation between attack scenarios, affected SDN layers or interfaces, and security. In contrast to Ruffy et al.'s approach, their threat analysis and proposed countermeasures are based entirely on a literature review \cite{ruffy-2016,iqbal-2019}.

Varadharajan et al. proposed a Policy-based Security Architecture (PbSA) for SDN. They grouped threats into four different categories that targeted different components of the SDN. Next, they identified five attack scenarios and formulated four security requirements to mitigate these threats. Furthermore, they developed a policy-based security application that is installed on the SDN controller's northbound interface to address these requirements. They assessed their security architecture's performance and its capacity to counter various threats and fulfil different security requirements. Since a single countermeasure cannot address all vulnerabilities, a suitable solution should include multiple mitigations to improve overall SDN security \cite{varadharajan-2019}.

Chica et al. performed a security assessment of SDN, emphasizing network threats, potential attack routes, and vulnerable areas across SDN layers and interfaces. They discussed the extent to which SDN can enhance network security and how the security of SDN can be improved. Furthermore, they identified possible weaknesses and discrepancies in SDN security mechanisms. Even though they outlined potential attack scenarios, they did not explore their technical implementation or how their impact could affect their SDN environment \cite{correa-chica-2020}.

Al-Saghier proposed an SDN framework capable of automatically launching known attacks against SDN components. This was accomplished by integrating additional elements into the SDN structure, namely the agent manager, application agent, agent channel, and agent host. The framework recognized misconfiguration, malware, and insider attacks as potential attack vectors. Al-Saghier also used a PoC testbed to measure the SDN's response to various attacks \cite{al-saghier-2019}.

Sjoholmsierchio et al. emphasized the enhancement of TLS encryption in OpenFlow to strengthen SDN security, noting that TLS is presently the sole protector of the control channel (according to them). They found that TLS was vulnerable to downgrading attacks and developed a protocol dialecting approach to protect the TLS-encrypted OpenFlow protocol. By extending protocol dialecting to a policy enforcement proxy, they also proposed an innovative policy-based networking approach. Furthermore, to demonstrate the effectiveness of their proposed security measures in detecting and preventing these attacks, they simulated a downgrade attack in a Mininet-emulated SDN environment. While their results showed an improvement in control channel security, it was at the cost of a 22\% increase in communication latency over standard TLS \cite{sjoholmsierchio-2021}.

Shaghaghi et al. discussed recent data plane mitigations, identifying five key SDN security features: centralised controller, open programmable interfaces, forwarding device management protocol, third-party network services, and virtualised logical networks. Their work underlined aspects of SDN that may potentially reveal security vulnerabilities, particularly in the data plane. They provided an overview on the security challenges present in the data plane, specified solution requirements, and presented existing resolutions. They also stressed the importance of having robust and flexible security measures in place to protect the SDN architecture from potential threats. Their analysis suggests that, despite advances in security mechanisms, SDN has still open security issues \cite{shaghaghi-2020}.

Jiasi et al. proposed a blockchain-based SDN security solution that decentralises the control plane without losing the overall network perspective. This mechanism, combined with fine-grained access control of network-wide resources and a secure controller-switch channel, guarantees authenticity, traceability, and accountability of application flows. They achieved this by introducing a blockchain layer between the control and data planes including attribute-based encryption for access control at the northbound interface. To summarise, the blockchain can be used to record all network flows and events, and even support secure protocol implementations using smart contracts \cite{weng-2019}.

Kreutz et al. analyzed numerous threat vectors that could exploit SDN vulnerabilities and presented a secure and reliable SDN controller platform design. They identified seven attack vectors and proposed various countermeasures. Their suggested platform design uses three SDN controllers, instead of one, to improve reliability through replication. The design also mandates that switches dynamically associate with multiple controllers. A key recommendation was to promote controller diversity and avoid relying on a single controller for replication. However, their SDN security and reliability approach of using multiple controllers is a known technique in DCNs and other large-scale networks \cite{kreutz-2013}.

Prathima Mabel et al. gave an overview of OpenFlow vulnerabilities including mitigations for securing the SDN controller, data plane, and OpenFlow channel.
They proposed the "flow tracer" solution designed to protect the SDN controller from misuse by identifying and isolating fraudulent flow entries potentially inserted by hackers. The flow tracer also employs symmetric key encryption for packets bound for the data plane. By simulating a DoS attack in the Mininet emulator, they validated their solution and demonstrated its effectiveness for improving SDN controller, data plane and OpenFlow channel security \cite{prathima-mabel-2019}.

Cabaj et al. outlined three primary SDN security concerns: OpenFlow protocol limitations, centralised network operations, and lack of middleboxes. They proposed a theoretical solution, the Distributed Frequent Sets Analyzer (DFSA), for detecting SDN threats using data mining in response to these concerns. DFSA is designed to detect quickly any attacks generating significant traffic. The solution integrates various modules into the SDN architecture to provide automatic responses upon detecting e.g., a DoS attack against the SDN controller \cite{cabaj-2014}.

Fawcett et al. presented a scalable network security framework for SDN using an ONOS controller-based, multi-level, distributed system that operates across the application, coordination, and collection layers. The framework, known as TENNISON, includes characteristics that make it efficient, proportionate, scalable, programmable, transparent, resilient, and interoperable. Furthermore, it enables network monitoring and remediation without disrupting other services and includes a security pipeline with four flow tables that take precedence over other network application tables. To demonstrate TENNISON's effectiveness, the authors built a PoC testbed and evaluated its performance and scalability against DoS, Distributed DoS (DDos), scanning and intrusion attacks \cite{fawcett-2018}.

\subsection{Summary}

As discussed in this section, the identified related work mainly focuses on conducting security analyses, implementing remediation, or introducing new security solutions. These works vary in their methodological approach, with some being empirically tested while others remain theoretical. A significant difference among the studies lies in the number of SDN layers/interfaces considered, with several only concentrating on specific parts of the SDN architecture. Unlike these works, this article provides a comprehensive analysis of threats and vulnerabilities using STRIDE and PASTA models, assesses risks in DC environments, simulates attack scenarios, and suggests methods to enhance SDN security. As a result, a comprehensive SDN Security Evaluation Framework is provided, which allows DC owners to evaluate their network security, understand potential implications, and act against complex security risks. The following table summarises the identified related work and compares it with this article: 
\vspace{-15pt}
\begin{table*}[!h]
	\caption{Summary of Literature Review on Evaluating Security in SDN (adapted from Ivki\'c et al., 2023).}
	\centering
	\resizebox{\textwidth}{!}{
		\begin{tabular}{|ccc|c|ccccc|l|}
			\hline
			 \multicolumn{3}{|c|}{\textbf{Security}}                                                                                                               & \multirow{2}{*}{\textbf{PoC}} & \multicolumn{5}{c}{\textbf{SDN Layer/Interface}} & \multicolumn{1}{|c|}{\multirow{2}{*}{\textbf{Sources}}} \\ \cline{1-3} \cline{5-9} 
			\multicolumn{1}{|c|}{Analysis}                & \multicolumn{1}{c|}{Remediation}             & \begin{tabular}[c]{@{}c@{}}New Security\\ Solution\end{tabular} &                               & \multicolumn{1}{c|}{App}                                         & \multicolumn{1}{c|}{App-Ctl}                                                           & \multicolumn{1}{c|}{Ctl}                                                               & \multicolumn{1}{c|}{Ctl-Data}                                                          & \multicolumn{1}{c|}{Data} &   \multicolumn{1}{c|}{}                                                            \\ \hline
			\multicolumn{1}{|c|}{\xmark}   & \multicolumn{1}{c|}{\cmark}   &                                                        &                               & \multicolumn{1}{c|}{\xmark \cmark} & \multicolumn{1}{c|}{\xmark \cmark}                       & \multicolumn{1}{c|}{\xmark \cmark}                       & \multicolumn{1}{c|}{\xmark \cmark}                       & \xmark \cmark &
			\textcolor{white}{a}\cite{ruffy-2016} \textit{Ruffy et al.}                       \\ \hline
			\multicolumn{1}{|c|}{(\xmark)} & \multicolumn{1}{c|}{(\cmark)} & \omark                                  &                               & \multicolumn{1}{c|}{\xmark \cmark} & \multicolumn{1}{c|}{\xmark \cmark \omark} & \multicolumn{1}{c|}{\xmark \cmark}                       & \multicolumn{1}{c|}{\xmark \cmark \omark} & \xmark \cmark  &
			\textcolor{white}{a}\cite{iqbal-2019} \textit{Iqbal et al.}                      \\ \hline
			\multicolumn{1}{|c|}{}                        & \multicolumn{1}{c|}{}                        & \omark                                  & \omark         & \multicolumn{1}{c|}{}                                            & \multicolumn{1}{c|}{}                                                                  & \multicolumn{1}{c|}{\omark}                                             & \multicolumn{1}{c|}{\omark}                                             & \omark   &
			\textcolor{white}{a}\cite{varadharajan-2019} \textit{Varadharajan et al.}                                           \\ \hline
			\multicolumn{1}{|c|}{(\xmark)} & \multicolumn{1}{c|}{(\cmark)} &                                                        &                               & \multicolumn{1}{c|}{\xmark \cmark} & \multicolumn{1}{c|}{\xmark \cmark}                       & \multicolumn{1}{c|}{\xmark \cmark}                       & \multicolumn{1}{c|}{\xmark \cmark}                       & \xmark \cmark     &
			\textcolor{white}{aa}\cite{correa-chica-2020} \textit{Chica et al.}                    \\ \hline
			\multicolumn{1}{|c|}{(\xmark)} & \multicolumn{1}{c|}{}                        &                                                        & \xmark         & \multicolumn{1}{c|}{\xmark}                       & \multicolumn{1}{c|}{}                                                                  & \multicolumn{1}{c|}{\xmark}                                             & \multicolumn{1}{c|}{}                                                                  & \xmark                 &
			\textcolor{white}{aa}\cite{al-saghier-2019} \textit{Al-Saghier}                             \\ \hline
			\multicolumn{1}{|c|}{}                        & \multicolumn{1}{c|}{}                        & \omark                                  & \omark         & \multicolumn{1}{c|}{}                                            & \multicolumn{1}{c|}{}                                                                  & \multicolumn{1}{c|}{}                                                                  & \multicolumn{1}{c|}{\omark}                                             & &
			\textcolor{white}{a}\cite{sjoholmsierchio-2021} \textit{Sjoholmsierchio et al.}                                                                   \\ \hline
			\multicolumn{1}{|c|}{(\xmark)} & \multicolumn{1}{c|}{(\cmark)} &                                                        &                               & \multicolumn{1}{c|}{\xmark}                       & \multicolumn{1}{c|}{\xmark}                                             & \multicolumn{1}{c|}{\xmark}                                             & \multicolumn{1}{c|}{\xmark}                                             & \xmark \cmark &
			\textcolor{white}{a}\cite{shaghaghi-2020} \textit{Shaghaghi et al.}                       \\ \hline
			\multicolumn{1}{|c|}{}                        & \multicolumn{1}{c|}{}                        & \omark                                  & \omark         & \multicolumn{1}{c|}{\omark}                       & \multicolumn{1}{c|}{\omark}                                             & \multicolumn{1}{c|}{\omark}                                             & \multicolumn{1}{c|}{\omark}                                             & \omark &
			\textcolor{white}{a}\cite{weng-2019} \textit{Jiasi et al.}                                              \\ \hline
			\multicolumn{1}{|c|}{(\xmark)} & \multicolumn{1}{c|}{(\cmark)} & \omark                                  &                               & \multicolumn{1}{c|}{\xmark \cmark} & \multicolumn{1}{c|}{\xmark \cmark}                       & \multicolumn{1}{c|}{\xmark \cmark \omark} & \multicolumn{1}{c|}{\xmark \cmark}                       & \xmark \cmark      &
			\textcolor{white}{a}\cite{kreutz-2013} \textit{Kreutz et al.}                  \\ \hline
			\multicolumn{1}{|c|}{(\xmark)} & \multicolumn{1}{c|}{(\cmark)} & \omark                                  & \omark         & \multicolumn{1}{c|}{}                                            & \multicolumn{1}{c|}{}                                                                  & \multicolumn{1}{c|}{\xmark \cmark \omark} & \multicolumn{1}{c|}{\xmark \cmark \omark} & \xmark \cmark \omark &
			\textcolor{white}{a}\cite{prathima-mabel-2019} \textit{Prathima Mabel et al.}  \\ \hline
			\multicolumn{1}{|c|}{}                        & \multicolumn{1}{c|}{}                        & \omark                                  &                               & \multicolumn{1}{c|}{}                                            & \multicolumn{1}{c|}{}                                                                  & \multicolumn{1}{c|}{\omark}                                             & \multicolumn{1}{c|}{}                                                                  & &
			\textcolor{white}{aa}\cite{cabaj-2014} \textit{Cabaj et al.}                                                                    \\ \hline
			\multicolumn{1}{|c|}{}                        & \multicolumn{1}{c|}{}                        & \omark                                  & \omark         & \multicolumn{1}{c|}{}                                            & \multicolumn{1}{c|}{}                                                                  & \multicolumn{1}{c|}{\omark}                                             & \multicolumn{1}{c|}{\omark}                                             & \omark &
			\textcolor{white}{aa}\cite{fawcett-2018} \textit{Fawcett et al.}                                             \\ \hline
			\multicolumn{1}{|c|}{\xmark}   & \multicolumn{1}{c|}{(\cmark)} &                                                        & \xmark         & \multicolumn{1}{c|}{\xmark \cmark}                                            & \multicolumn{1}{c|}{\xmark \cmark}                       & \multicolumn{1}{c|}{\xmark \cmark}                       & \multicolumn{1}{c|}{\xmark \cmark}                       & \xmark \cmark &
			\textcolor{white}{a}\textbf{This Article}                      \\ \hline
			\multicolumn{10}{l}{}\\
			\multicolumn{10}{l}{\textcolor{white}{aaaaa}\xmark\textcolor{white}{a}\begin{tabular}[c]{@{}l@{}}Security\\ Analysis\end{tabular}\textcolor{white}{aaaaaaa}\cmark\textcolor{white}{a}\begin{tabular}[c]{@{}l@{}}Security\\ Remediation\end{tabular}\textcolor{white}{aaaaaaa}\omark\textcolor{white}{a}\begin{tabular}[c]{@{}l@{}}New Security\\ Solution\end{tabular}\textcolor{white}{aaaaaaa}()\textcolor{white}{a}\begin{tabular}[c]{@{}l@{}} Fully Based on \\ Literature Review\end{tabular}}
		\end{tabular}
	}
	\label{related_work_table}
\end{table*}

\vspace{-15pt}
As shown in Table \ref{related_work_table}, only five papers comprehensively investigated all SDN components. Jiasi et al. \cite{weng-2019} proposed a solution that can secure the entire SDN architecture, while most other solutions focus on the control and data planes and their communication channel. Ruffy et al. \cite{ruffy-2016} employed the STRIDE model for SDN security analysis, unlike most papers that primarily used literature reviews. Five of eight papers validated their proposed security solutions using a PoC.

Compared to other research, this article utilizes two security models, STRIDE and PASTA, to analyze SDN threats and vulnerabilities across all components. The combination of these two models enables a finer-tuned security evaluation, resulting in tailored security measures for SDN. Furthermore, it includes a risk and impact assessment based on CVSS v3.1, which ranks threats by severity and describes their effects on DC environments. After identifying and analysing SDN security threats, the next step is to validate them in an experimental study. Therefore, an SDN testbed setup is built in Mininet to outline the technical implementation of the attacks on one hand, and verify the previously analysed impact of the threats on the other hand. Although Al-Saghier \cite{al-saghier-2019} validated security analysis in a test setup, he neither did elaborate on the technical aspects of the attack scenario nor did he consider the entire SDN architecture in his PoC.

\vspace{-15pt}
\section{Framework and Testbed}
\vspace{-10pt}

The main beneficiaries of the emerging architectural transformation sparked by SDN are distributed cloud applications hosted in DC environments. From a security standpoint, however, SDN brings new challenges, as a centralised network controller can be both a strength and a vulnerability. To be more precise, as soon as the SDN controller or the application is compromised, the entire network is affected \cite{al-saghier-2019}. To gain a better understanding of the potential threats and vulnerabilities of SDN components in DC environments a Security Evaluation Framework and its four phases are presented in this section. In this regard, we explain how the framework is used for identifying threats and vulnerability, analysing risks and their impact on DC environments, modelling possible attack scenarios, and applying mitigation measures. Furthermore, we present our experimental testbed that simulates basic SDN functionalities in DCs. The following figure shows the proposed framework:
\vspace{-15pt}
\label{sec:framework}
\begin{figure*}[!h]
	\centering
	\includegraphics[width=\textwidth]{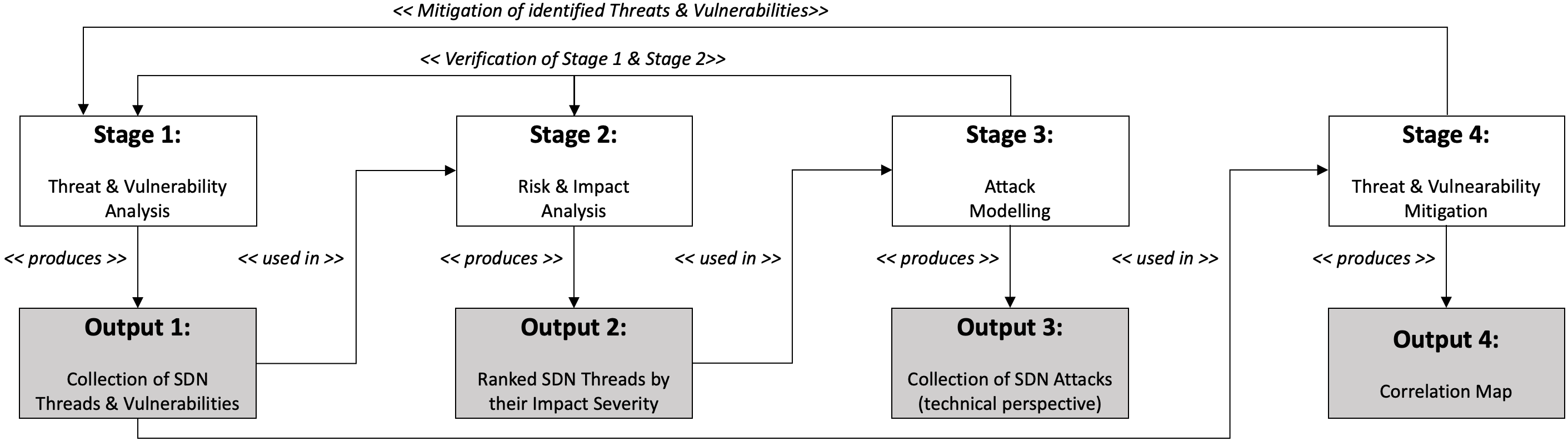}
	\caption{Security Evaluation Framework for SDN Architectures in DC Environments (Ivki\'c et al., 2023).}
	\label{fig:fig1}
\end{figure*}

\vspace{-25pt}
\subsection{Stage 1 – Threat \& Vulnerability Analysis}

Ruffy et al suggest performing a risk-benefit analysis when considering the deployment of SDN technology, as an SDN controller represents a single point of failure for the entire network. However, prior to any risk assessment, the first step must be the identification of threats. Threat groups or categories are typically used because many threats can be relevant to a security assessment. Microsoft's STRIDE model is a common approach to technically distinguishing security design flaws by applying a generic set of threats based on their names, an acronym derived from the initials of the six main threat categories: Spoofing, Tampering, Repudiation, Information Disclosure, DoS and Elevation of Privilege \cite{hewko-2021,ruffy-2016,shevchenko-2018}.

\begin{figure}[!h]
	\centering
	\includegraphics[width=12.1cm]{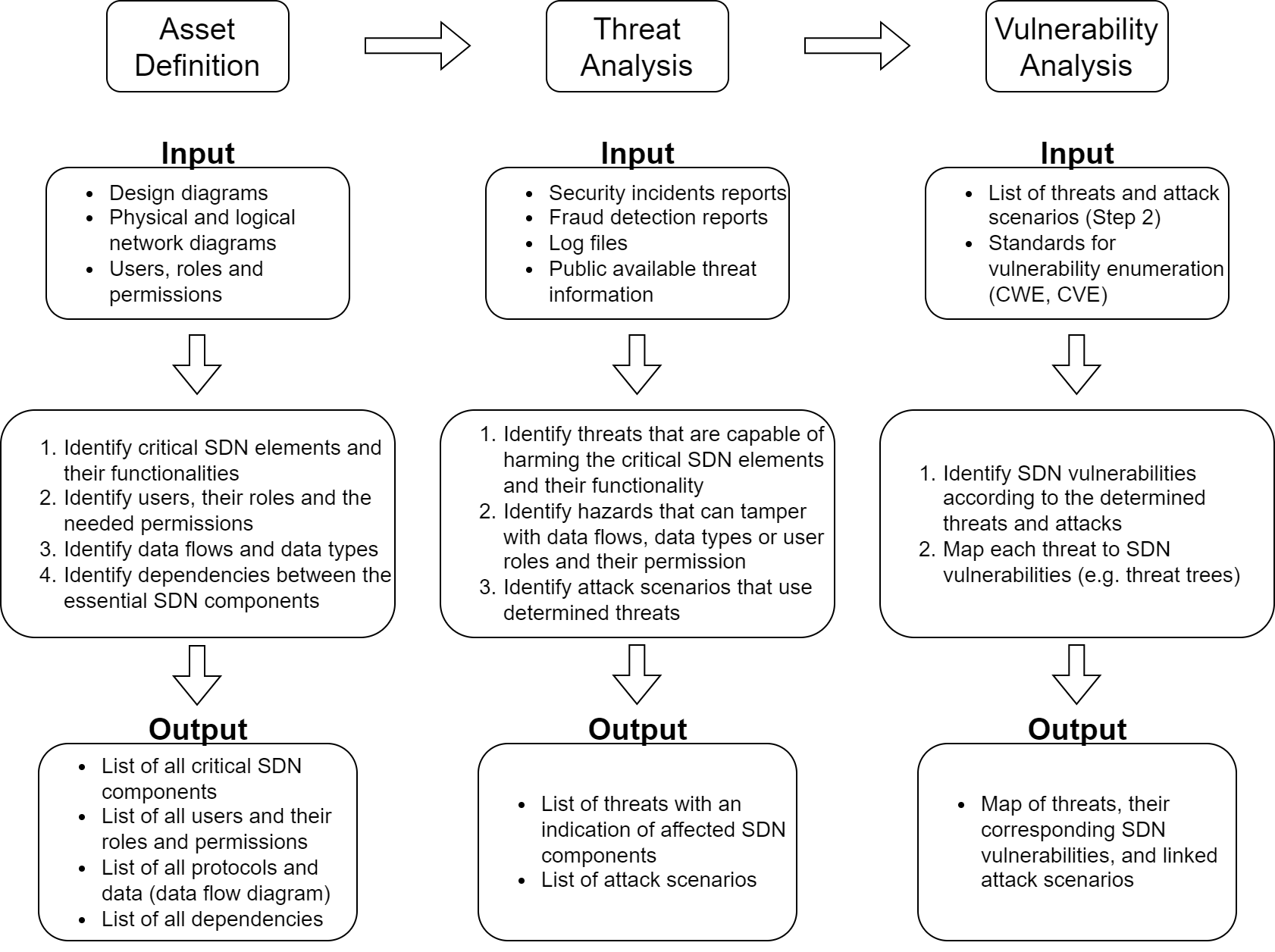}
	\caption{PASTA Model Adaption for the Threat and Vulnerability Analysis (adapted from Verspriter, 2022).}
	\label{adapted_pasta}
\end{figure}

Threat categorisation using STRIDE provides a useful overview of the relevance of certain threat groups that are more generally applicable to any network or application. Therefore, the risk-based threat modelling methodology called PASTA is used in addition to model SDN-specific threats and assess their business impact. However, the PASTA model has more steps than can be applied to a comprehensive analysis of the security of SDN. Consequently, the original seven-step PASTA model is adapted into a three-step process for SDN threat and vulnerability analysis, with modified action points within each step, as shown in Figure \ref{adapted_pasta}. In summary, both the STRIDE and PASTA approaches to SDN-based security threat and vulnerability assessment are incorporated into the first step of the Security Evaluation Framework.

\vspace{-15pt}

\vspace{-25pt}
\subsection{Stage 2 – Risk \& Impact Analysis}

After the first step, where the SDN threats and vulnerabilities have been identified, the next step is to categorize them based on their severity impact on DC environments. In order to speed up the risk and impact analysis, threats with similar impacts are then grouped into Threat Categories (TCs). This categorisation is based on a comparison of the results of the STRIDE and PASTA analyses in order to eliminate possible redundancies. Compared to the PASTA analysis, this step of defining TCs is done in reverse order, by first calculating the impact of vulnerabilities on the SDN before mapping them to threats. 

The next step is to perform a risk and impact analysis using the CVSS calculator provided by the National Vulnerability Database (NVD). The CVSS v3.1 scoring method uses three different sets of metrics to calculate a base score and an overall score for each identified threat. The base score is the basis for determining the severity. The total score represents the severity considering the environmental metric group.
The basic metric group defines the intrinsic characteristics of a vulnerability that are constant over time and across environments. The temporal metric group represents the aspects of a vulnerability that may change over time but not across environments. The environmental metric group describes vulnerability factors relevant and unique to a particular environment. 

The TC has a greater impact on the environment than assumed by the threat assessment if the total score is higher than the base score, and vice versa. A CVSS score is also represented as a vector string, a compressed textual representation of the values used to derive the score. In summary, the severity and impact of the previously identified threats and vulnerabilities on SDN-based DC environments are calculated in the second stage of the Security Evaluation Framework using a standardised approach (CVSS v3.1). The following figure shows how the CVSS score is calculated and assigned on a scale between 0 and 10:

\vspace{-15pt}
\begin{figure}[!h]
	\centering
	\includegraphics[width=\textwidth]{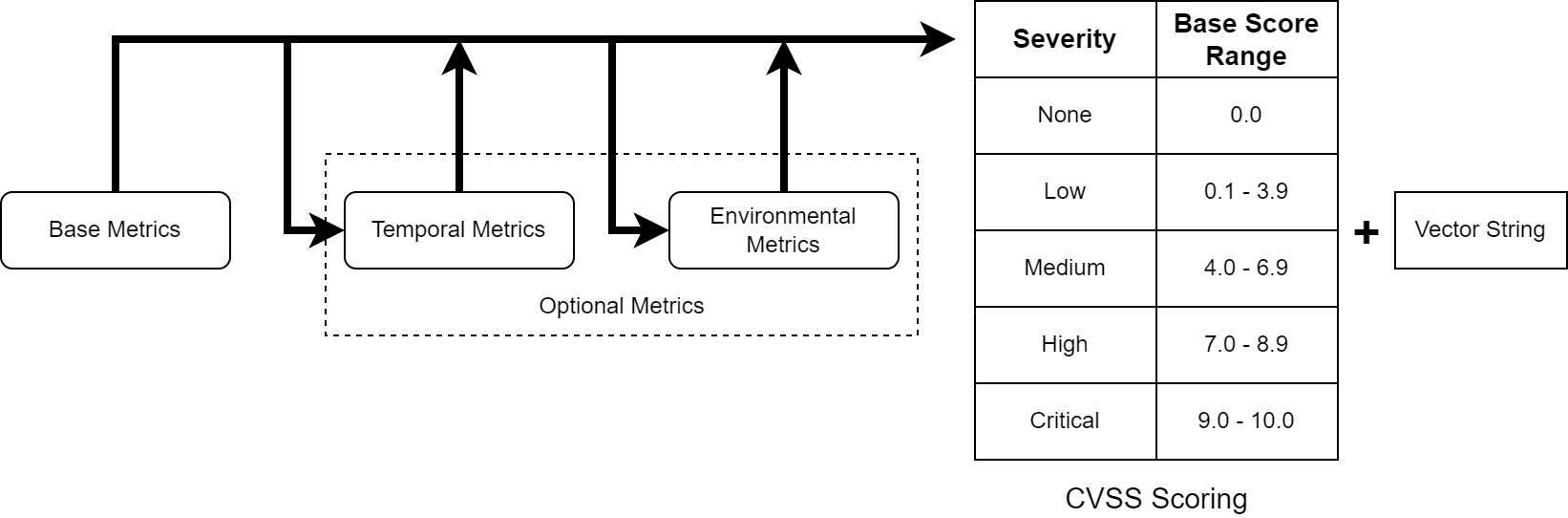}
	\caption{CVSS Scoring Flow with Severity Base Score Mapping (adapted from First, 2019) }
\label{cvss_scoring}
\end{figure}

\vspace{-35pt}
\subsection{Stage 3 – Attack Modelling}

The third phase focuses on creating attack scenarios to validate their impact against an SDN testbed once threats and vulnerabilities have been identified and a risk and impact analysis has been completed. By modelling attacks for each TC, the overall goal of this phase is to validate the results of the risk and impact analysis. This can be done either on a real SDN-based DCN or by creating a realistic test environment (testbed) where the attacks can be executed without disrupting the production environment. Furthermore, based on the calculated CVSS score, it may be useful in some cases to select the TCs to be tested. For example, modelling is done first for the highest probability and highest severity threats, and then for the lower probability and lower severity threats. In summary, the third step proposes to conduct an experimental study where attacks are modelled as an additional measure to validate the ranked risks and their severity impact from the previous step.

\vspace{-10pt}
\subsection{Stage 4 – Threat \& Vulnerability Mitigation}

Countermeasures that can be applied to mitigate the identified threats and thereby improve the overall security of SDN are provided in the final stage of the framework. In general, SDN countermeasures either include methods to mitigate a specific threat (e.g., man-in-the-middle attacks), or represent central solutions that mitigate multiple threats (e.g., PbSA). The first category of mitigations can be mapped directly to the SDN threats identified in Stage 1. This assists users of this framework to apply the necessary measures to reduce the identified threats and weaknesses of their SDN. Such countermeasures usually do not require complex implementation of external tools or software. Instead, the underlying operating system (OS) or protocols used in SDN environments often provide security enhancements but are often not enabled by default.

A more centralised solution is provided by the second category of mitigations. Rather than applying countermeasures to each identified threat, a more generic solution can be implemented to secure multiple SDN components at once against different types of attacks. These solutions typically involve complex and time-consuming external systems that need to be integrated into the existing infrastructure. On the one hand, this approach allows multiple threats to be prevented with a single solution. However, on the other hand, the implementation of a central solution requires external systems that have the potential to introduce new vulnerabilities into the SDN. In conclusion, the final stage of the framework provides mitigations that can either be directly mapped to a threat or introduced as a central solution to mitigate multiple threats.

\vspace{-10pt}
\subsection{Experimental Testbed}

A testbed that simulates the basic functionality of an SDN in DC environments and follows a simple architecture to reduce any unnecessary functionality overhead has been implemented for the PoC evaluation. To achieve the right balance of simplicity and functionality, Mininet is used to create a virtual SDN that takes advantage of the process-based virtualisation and network namespace capabilities of the latest Linux kernel. Mininet simulates the links as virtual Ethernet pairs, which are in the Linux kernel, and which connect the switches to the hosts. A key feature for SDN simulation is the software-based Open vSwitch (OVS) that uses the OpenFlow controller to communicate with the SDN controller \cite{mininet-2022}. 

Although Mininet can emulate a rudimentary SDN controller, it does not provide sufficient functionality for a test network that simulates multi-tenancy in a DC environment. Therefore, the ONOS SDN controller is used as an additional external controller to manage the switches hosted by Mininet. This combination of the emulated network hosted by Mininet and the ONOS SDN controller provides a suitable testbed architecture for the PoC evaluation.

The architecture of the test network has been designed to be as simple as possible, yet complex enough to measure the impact of different attacks. The testbed network topology consists of an SDN controller connected to three switches, each connecting three VMs. Rather than running on a separate machine, the ONOS controller is a platform where SDN applications can be installed directly \cite{onos-2014}. One of these applications, the Virtual Private LAN Service (VPLS), was used in the testbed for isolating multiple domains and simulating multi-tenancy \cite{lasserre-2007}.

The testbed has been configured so that hosts can only reach other hosts from the same VPLS. This service will be used later in the testbed to verify whether certain attack scenarios can bypass the isolation or reconfigure the service to allow communication between isolated hosts. An additional node running Kali Linux is used to run various attack scenarios (Stage 3 of the Security Evaluation Framework) against the testbed. Kali Linux is a Debian-based open source distribution for advanced penetration testing. The Kali VM provides a set of tools for testing security in different categories. These include penetration testing, security research, computer forensics and reverse engineering. Due to the comprehensive set of security tools, Kali is a suitable source for the attack of the SDN-based testbed. The figure below shows the described testbed:

\vspace{-20pt}
\begin{figure*}[!h]
	\centering
	\includegraphics[width=\textwidth]{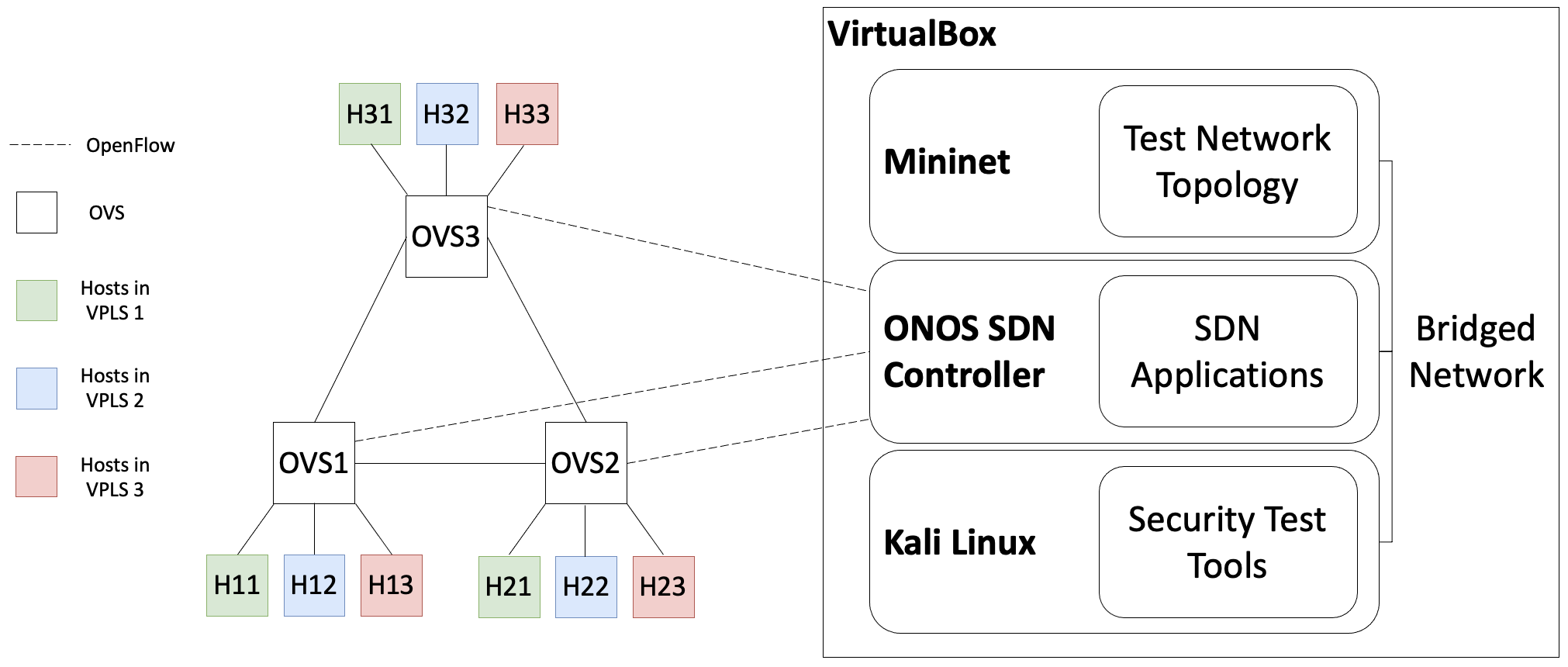}
	\caption{SDN Testbed using Mininet, ONOS Controller and VPLS Services for Network Isolation (Ivki\'c et al., 2023).}
	\label{fig:fig2}
\end{figure*}

\vspace{-30pt}
\section{Threat \& Vulnerability Analysis (Stage 1)}
\label{sec:Stage1}
\vspace{-5pt}

As shown in Figure \ref{fig:fig1}, the first stage of the Security Evaluation Framework suggests performing an SDN threat and vulnerability analysis using STRIDE and PASTA. Initially, the primary SDN components were separately evaluated for all six threats of the STRIDE analysis. Next, the Threat Modelling Tool from Microsoft was used to execute a model-driven STRIDE analysis. The tool's input is a basic SDN model comprising an SDN application, two controllers (typical in DC environments), and one forwarding device.

The tool can execute a STRIDE analysis per element, meaning it is unnecessary to model each SDN component separately. Furthermore, it is possible to create trust boundaries and configure security parameters per component/data flow. Both adjustment options are not considered to generate the maximum possible number of vulnerability suggestions from the tool. Vulnerabilities that are not applicable to an SDN network or are redundant are rejected afterward.

The mind map from Figure \ref{fig:mindmap} represents the baseline of vulnerabilities which is then again evaluated using PASTA to identify additional threats and vulnerabilities that might have been missed with only the STRIDE approach. The input information for that step can include security and fraud reports from an existing SDN network to adapt the threat analysis according to previous reports. One of the most comprehensive sources of such information is the MITRE ATT\&CK database, which provides over 200 attack techniques categorized in different classes and linked with mitigations \cite{mitre-corporation-2021}. The following mind map summarizes all identified SDN security threats and vulnerabilities using STRIDE:

\vspace{-20pt}
\begin{figure*}[!h]
	\centering
	\includegraphics[width=\textwidth]{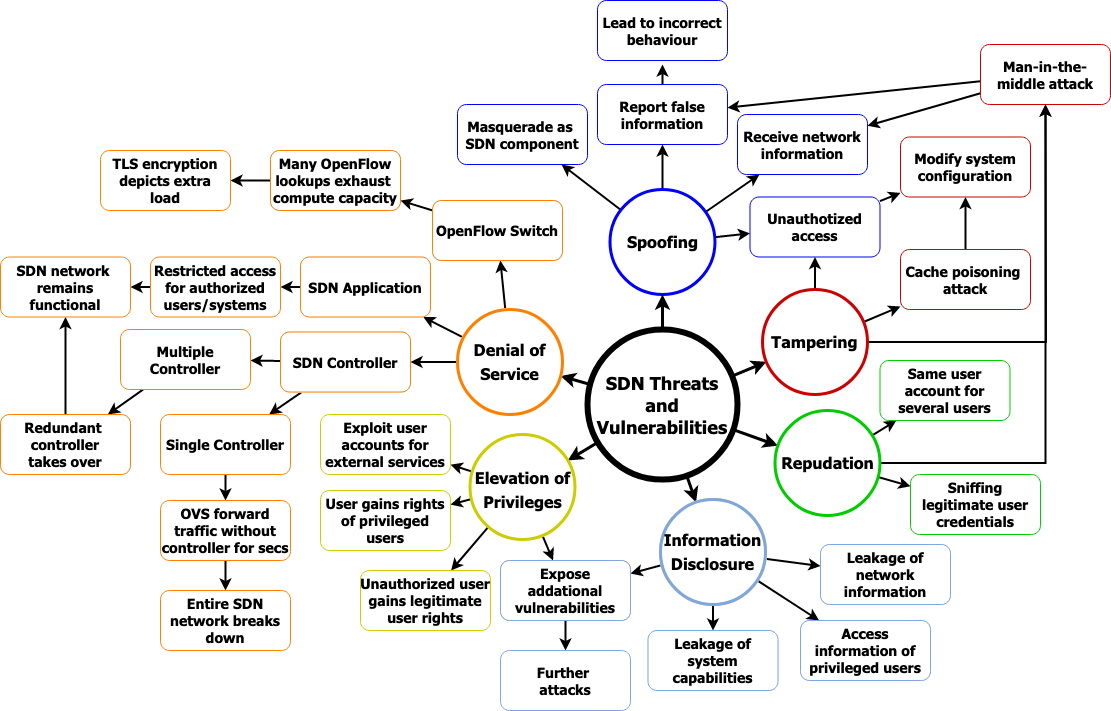}
	\caption{Mind Map of identified SDN Threats and Vulnerabilities using STRIDE.}
	\label{fig:mindmap}
\end{figure*}

\vspace{-15pt}
The following table shows the additionally identified threats using PASTA based on the MITRE ATT\&CK database: 
\vspace{-15pt}
\begin{table*}[!h]
	\caption{PASTA Threat Analysis based on MITRE Corporation (adapted from Ivki\'c et al., 2023).}
	\centering
	\resizebox{\textwidth}{!}{
		\begin{tabular}{|ll|l|}
			\hline
			\multicolumn{2}{|c|}{\textbf{Threat}}                                                                  & \multicolumn{1}{c|}{\textbf{Description}}                                                                                                                                                                                                                                                                                                                                                                                              \\ \hline
			\multicolumn{1}{|l|}{T1} & \begin{tabular}[c]{@{}l@{}}Command and\\ Scripting Interpreter\end{tabular} & \begin{tabular}[c]{@{}l@{}}\tabitem adversaries may abuse command and script interpreters to execute command, scripts, or binaries\\ \tabitem attackers may execute commands via terminals/shells or use remote services to achieve remote execution\end{tabular}                                                                                                                                        \\ \hline
			\multicolumn{1}{|l|}{T2} & \begin{tabular}[c]{@{}l@{}}Modify\\ Authentication Process\end{tabular}     & \begin{tabular}[c]{@{}l@{}}\tabitem intruders modify the authentication mechanism\\ \tabitem access user credentials or enable unwarranted access to accounts\\ \tabitem compromised credentials used to bypass access controls placed on various systems and may even be used for persistent \\ access to remote systems and externally available services\end{tabular}                  \\ \hline
			\multicolumn{1}{|l|}{T3} & Impair Defenses                                                             & \begin{tabular}[c]{@{}l@{}}\tabitem maliciously modify components to hinder or disable defensive mechanism\\ \tabitem e.g. modify/disable the Access Control List (ACL) for the overlay traffic on the SDN application\end{tabular}                                                                                                                                                                      \\ \hline
			\multicolumn{1}{|l|}{T4} & \begin{tabular}[c]{@{}l@{}}Network Boundary\\ Bridging\end{tabular}         & \begin{tabular}[c]{@{}l@{}}\tabitem attackers may bridge network boundaries by compromising SDN application, controller, or forwarding device\\ \tabitem bypass traffic isolation that separates tenant networks\\ \tabitem enable movement into new victim environments\end{tabular}                                                                                                     \\ \hline
			\multicolumn{1}{|l|}{T5} & Weaken Encryption                                                           & \begin{tabular}[c]{@{}l@{}}\tabitem compromising network encryption capabilities to bypass encryption that would otherwise protect data communication\\ \tabitem can be achieved by behaviours such as modifying system image, reducing key space, and disabling crypto hardware\\ \tabitem greater risk of unauthorized disclosure and data manipulation\end{tabular}                    \\ \hline
			\multicolumn{1}{|l|}{T6} & Network Sniffing                                                            & \begin{tabular}[c]{@{}l@{}}\tabitem sniffing network traffic to capture information about the SDN environment, running services, authentication, etc.\\ \tabitem adversaries may set an interface to promiscuous mode to passively access data in transit\end{tabular}                                                                                                                                   \\ \hline
			\multicolumn{1}{|l|}{T7} & Automated Exfiltration                                                      & \begin{tabular}[c]{@{}l@{}}\tabitem leverage traffic mirroring to automate data exfiltration over compromised networks\\ \tabitem traffic mirroring is a native feature to forward duplicated traffic to one or more destinations for analyzing\\ \tabitem adversaries may use traffic duplication in conjunction with network sniffing, input capture, or man-in-the-middle\end{tabular} \\ \hline
			\multicolumn{1}{|l|}{T8} & Active Scanning                                                             & \begin{tabular}[c]{@{}l@{}}\tabitem executing reconnaissance scans to gather network information via native features of network protocols such as\\ Internet Control Message Protocol (ICMP)\\ \tabitem information from those scans may expose opportunities for other attacks\end{tabular}                                                                                                             \\ \hline
		\end{tabular}%
	}
	\label{pasta_threats}
\end{table*}

\vspace{-15pt}
The Open Web Application Security Project (OWASP) Top 10 list from 2021 serves as the basis for the in-depth threat analysis for the application layer. The following table summarises additional identified SDN threats based on OWASP:
\vspace{-25pt}
\begin{table*}[!h]
	\caption{PASTA Threat Analysis based on OWASP Top 10 (2021) for the SDN Application Layer (adapted from Ivki\'c et al., 2023).}
	\centering
	\resizebox{\columnwidth}{!}{%
		\begin{tabular}{|cc|l|}
			\hline
			\multicolumn{2}{|c|}{\textbf{Threat}}                                                                            & \multicolumn{1}{c|}{\textbf{Description}}                                                                                                                                                                                                                                                                                                                                                                                                                                   \\ \hline
			\multicolumn{1}{|c|}{T9}  & \begin{tabular}[c]{@{}c@{}}Broken Access \\ Control\end{tabular}                     & \begin{tabular}[c]{@{}l@{}}\tabitem access control enforces policies to ensure that users cannot act outside of their intended permissions\\ \tabitem failures lead to unauthorized information disclosure, modification, or destruction of all data\\ \tabitem function execution outside the user its limits\end{tabular}                                                                                                    \\ \hline
			\multicolumn{1}{|c|}{T10} & \begin{tabular}[c]{@{}c@{}}Cryptographic \\ Failures\end{tabular}                    & \begin{tabular}[c]{@{}l@{}}\tabitem identify the protection needs of data in transit\\ \tabitem for sensitive data, robust encryption mechanisms need to be in place\end{tabular}                                                                                                                                                                                                                                                             \\ \hline
			\multicolumn{1}{|c|}{T11} & Injection                                                                            & \begin{tabular}[c]{@{}l@{}}\tabitem typical injections are Structured Query Language (SQL), NoSQL or OS command injection\\ \tabitem supplied data is not validated by the application\\ \tabitem hostile data is used to extract sensitive records\end{tabular}                                                                                                                                                               \\ \hline
			\multicolumn{1}{|c|}{T12} & Insecure Design                                                                      & \begin{tabular}[c]{@{}l@{}}\tabitem broad category representing different weaknesses\\ \tabitem missing or ineffective control design\\ \tabitem differentiate between design flaws and implementation defects because of different root causes and\\ remediations\\ \tabitem a secure design can still have implementation defects, but a perfect implementation cannot fix an \\ insecure design\end{tabular} \\ \hline
			\multicolumn{1}{|c|}{T13} & \begin{tabular}[c]{@{}c@{}}Security \\ Misconfiguration\end{tabular}                 & \tabitem repeatable application security configuration process enables system robustness                                                                                                                                                                                                                                                                                                                                                                     \\ \hline
			\multicolumn{1}{|c|}{T14} & \begin{tabular}[c]{@{}c@{}}Vulnerable and\\ Outdated Components\end{tabular}         & \begin{tabular}[c]{@{}l@{}}\tabitem keep an overview of the versions of all components\\ \tabitem check software version vulnerabilities\\ \tabitem regularly vulnerability scans\end{tabular}                                                                                                                                                                                                                                 \\ \hline
			\multicolumn{1}{|c|}{T15} & \begin{tabular}[c]{@{}c@{}}Identification and\\ Authentication Failures\end{tabular} & \begin{tabular}[c]{@{}l@{}}\tabitem validate user its identity, authentication, and session management is critical to protect against\\ authentication-related attacks\end{tabular}                                                                                                                                                                                                                                                                          \\ \hline
			\multicolumn{1}{|c|}{T16} & \begin{tabular}[c]{@{}c@{}}Software and Data\\ Integrity Failures\end{tabular}       & \begin{tabular}[c]{@{}l@{}}\tabitem code and infrastructure that does not protect against integrity violations\\ \tabitem application may rely on untrusted sources, repositories, or libraries\\ \tabitem attacker could upload their own updates to distribute malicious data across several systems\end{tabular}                                                                                                            \\ \hline
			\multicolumn{1}{|c|}{T17} & \begin{tabular}[c]{@{}c@{}}Security Logging and\\ Monitoring Failures\end{tabular}   & \begin{tabular}[c]{@{}l@{}}\tabitem without logging and monitoring, breaches remain undetected\\ \tabitem systems need to detect, escalate, and respond to active breaches\end{tabular}                                                                                                                                                                                                                                                       \\ \hline
			\multicolumn{1}{|c|}{T18} & \begin{tabular}[c]{@{}c@{}}Server-Side Request\\ Forgery (SSRF)\end{tabular}         & \begin{tabular}[c]{@{}l@{}}\tabitem occurs whenever a web application is fetching a remote resource without validating the \\ user-supplied Uniform Resource Locator (URL)\\ \tabitem enables an adversary to coerce the application to send a crafted request to an unexpected destination\end{tabular}                                                                                                                                      \\ \hline
		\end{tabular}%
	}
	\label{owasp_top}
\end{table*}

\vspace{-20pt}
Both tables show that a threat analysis solely based on STRIDE would be insufficient since PASTA identified additional threats that affect all SDN components. The two tables on the next page summarise the last step of the PASTA framework including vulnerabilities based on the previously identified threats from (1) the MITRE Corporation (Table \ref{vulnerability_mitre}), and (2) OWASP Top 10 (Table \ref{vulnerability_owasp}).

In summary, Stage 1 analysed the SDN threats and vulnerabilities based on the STRIDE and PASTA models. The main SDN components were investigated separately for all six threats of the STRIDE analysis. Additionally, the Microsoft Threat Modelling Tool was used to verify the outcomes of the STRIDE assessment. The outcome of the STRIDE model, shown in Figure \ref{fig:mindmap}, provides a generic map of SDN threats and vulnerabilities. It gives an overview of the six common threats that the STRIDE model covers, and the associated vulnerabilities and effects on the SDN architecture. However, the STRIDE approach does not include additional threats that may be relevant for a comprehensive analysis. Therefore, the PASTA model was adapted and used to execute an additional SDN threat and vulnerability analysis. The results of both approaches confirmed that it is necessary to combine them to identify as many SDN security flaws as possible.

\begin{table*}[!h]
	\caption{Vulnerability Analysis based on MITRE Corporation (Ivki\'c et al., 2023).}
	\centering
	\resizebox{\columnwidth}{!}{%
		\begin{tabular}{|c|l|}
			\hline
			\textbf{Vulnerabilities} & \multicolumn{1}{c|}{\textbf{Description}}                                                                                                                                                                                                                                                                                                                                                                               \\ \hline
			V1                     & \begin{tabular}[c]{@{}l@{}}\tabitem permission is granted for non-signed scripts\\ \tabitem no execution prevention via application control mechanisms\\ \tabitem presence of unnecessary or unused shells or interpreters\end{tabular}                                                                                                                                    \\ \hline
			V2                     & \begin{tabular}[c]{@{}l@{}}\tabitem lack of multi-factor authentication\\ \tabitem lack of privileged account management - ignore the least privilege principle\\ \tabitem lack of restricted file and directory permissions\end{tabular}                                                                                                                                  \\ \hline
			V3                     & \begin{tabular}[c]{@{}l@{}}\tabitem lack of restricted file and directory permissions\\ \tabitem lack of user account management - wrong permissions for user accounts\end{tabular}                                                                                                                                                                                                       \\ \hline
			V4                     & \begin{tabular}[c]{@{}l@{}}\tabitem lack of credential access protection - local passwords may stored in plain-text\\ \tabitem lack of multi-factor authentication\\ \tabitem weak password policies\\ \tabitem lack of privileged account management - ignore least privilege principle or using the same credentials\\ across multiple systems\end{tabular} \\ \hline
			V5                     & \tabitem since the threat is based on the abuse of system features, there is no vulnerability that can easily mapped                                                                                                                                                                                                                                                                                     \\ \hline
			V6                     & \begin{tabular}[c]{@{}l@{}}\tabitem lack of multi-factor authentication\\ \tabitem lack of data encryption\end{tabular}                                                                                                                                                                                                                                                                   \\ \hline
			V7                     & \tabitem lack of data encryption                                                                                                                                                                                                                                                                                                                                                                         \\ \hline
			V8                     & \tabitem lack of traffic monitoring                                                                                                                                                                                                                                                                                                                                                                      \\ \hline
		\end{tabular}%
	}
	\label{vulnerability_mitre}
\end{table*}

\vspace{-15pt}
\begin{table*}[!h]
	\caption{Vulnerability Analysis based on OWASP Top 10 (Ivki\'c et al., 2023).}
	\centering
	\resizebox{\columnwidth}{!}{%
		\begin{tabular}{|c|l|}
			\hline
			\textbf{Vulnerabilities} & \multicolumn{1}{c|}{\textbf{Description}}                                                                                                                                                                                                                                                                                                                                                                                                                                                                                                                                                                                                                                        \\ \hline
			V9                       & \begin{tabular}[c]{@{}l@{}}\tabitem violation of principle of least privilege or deny by default\\ \tabitem bypassing access control checks by modifying the URL\\ \tabitem accessing API with missing access controls\\ \tabitem permitting viewing or editing someone else its account, by providing its unique identifier\end{tabular}                                                                                                                                                                                                                                                                            \\ \hline
			V10                      & \begin{tabular}[c]{@{}l@{}}\tabitem data transmitted in clear-text by using protocols like HTTP, SMTP, File Transfer Protocol (FTP)\\ \tabitem old or weak cryptographic algorithms or protocols\\ \tabitem encryption not enforced\\ \tabitem lack of validation of received server certificate and the trust chain\\ \tabitem usage of deprecated hash functions such as Message-Digest Algorithm 5 (MD5) or \\ Secure Hash Algorithm 1 (SHA-1)\end{tabular}                                                                                                                                        \\ \hline
			V11                      & \begin{tabular}[c]{@{}l@{}}\tabitem data transmitted in clear-text by using protocols like HTTP, SMTP, File Transfer Protocol (FTP)\\ \tabitem user-supplied data is not validated, filtered, or sanitized by the application\\ \tabitem dynamic queries are used directly in the interpreter\end{tabular}                                                                                                                                                                                                                                                                                                                          \\ \hline
			V12                      & \begin{tabular}[c]{@{}l@{}}\tabitem lack of business risk profiling\\ \tabitem failure to determine what level of security design is required\end{tabular}                                                                                                                                                                                                                                                                                                                                                                                                                                                                                         \\ \hline
			V13                      & \begin{tabular}[c]{@{}l@{}}\tabitem missing appropriate security hardening\\ \tabitem unnecessary features are enabled or installed\\ \tabitem default accounts and their passwords are still enabled and unchanged\\ \tabitem the latest security features are disabled or not configured\\ \tabitem the software is out of date or vulnerable\end{tabular}                                                                                                                                                                                                                                          \\ \hline
			V14                      & \begin{tabular}[c]{@{}l@{}}\tabitem no overview of the versions of all components\\ \tabitem software is vulnerable, unsupported, or out of date\\ \tabitem lack of regularly vulnerability scans\\ \tabitem lack of upgrading the underlying platform\\ \tabitem lack of compatibility tests after updated, upgraded, or patched libraries\\ \tabitem security misconfiguration on components (as mentioned in Security Misconfiguration)\end{tabular}                                                                                                                                \\ \hline
			V15                      & \begin{tabular}[c]{@{}l@{}}\tabitem application permits brute force or other automated attacks\\ \tabitem usage of default, weak, or well-known passwords, such as "admin/admin"\\ \tabitem usage of weak or ineffective credential recovery and forgot-password processes\\ \tabitem usage of plain text or weakly hashed passwords data stores (as mentioned in Cryptographic Failures)\\ \tabitem missing multi-factor authentication\\ \tabitem exposition of session identifier in the URL\\ \tabitem reuse session identifier after successful login\end{tabular} \\ \hline
			V16                      & \begin{tabular}[c]{@{}l@{}}\tabitem application relies upon plugins, libraries, or modules from untrusted sources or repositories\\ \tabitem lack of sufficient integrity verification for auto-update functionality\end{tabular}                                                                                                                                                                                                                                                                                                                                                                                                                  \\ \hline
			V17                      & \begin{tabular}[c]{@{}l@{}}\tabitem lack of logging auditable events, such as logins and failed logins\\ \tabitem unclear log messages\\ \tabitem logs of applications and APIs are not monitored for suspicious activity\\ \tabitem logs are only stored locally\\ \tabitem alerting thresholds and response escalation processes are not in place\\ \tabitem application cannot detect, escalate, or alert for active attacks in real-time or near real-time\end{tabular}                                                                                                            \\ \hline
			V18                      & \begin{tabular}[c]{@{}l@{}}\tabitem application is fetching a remote resource without validating the user-supplied URL\end{tabular}                                                                                                                                                                                                                                                                                                                                                                                                                                                                                                                             \\ \hline
		\end{tabular}%
	}
	\label{vulnerability_owasp}
\end{table*}
\newpage
\section{Risk \& Impact Analysis (Stage 2)}
\label{sec:Stage2}
\vspace{-5pt}

The second stage of the Security Evaluation Framework employs the threats and vulnerabilities identified in Stage 1 (Output 1) as input for a risk and impact analysis. The initial step groups the identified security threats from both assessments (STRIDE and PASTA) into categories with similar effects on the SDN network. It is crucial to consolidate the identified threats to allow for a more efficient risk and impact analysis. Grouping threats with similar impacts on the DC environment facilitates faster assessments through consistent comprehension. To achieve such classification, the outcomes from the STRIDE and PASTA analyses are compared initially to eliminate potential redundancies. As such, the six threats inherited by STRIDE are scrutinized against the root threats of the PASTA assessment. For instance, "Elevation of Privileges" from STRIDE is closely connected to "Broken Access Control" from PASTA.

In the second step, it is necessary to differentiate the threats based on their severity. It makes sense to distinguish two identical vulnerabilities if their impacts on the SDN are different. For example, from a security perspective, an intruder acquiring tenant permissions on the SDN application is significantly different from an intruder gaining admin permissions.

\vspace{-15pt}
\begin{table*}[!h]
	\caption{CVSS Scores and Severity of TCs (adapted from Ivki\'c et al., 2023).}
	\centering
	\resizebox{\textwidth}{!}{%
		\begin{tabular}{|c|cl|c|c|c|}
			\hline
			\textbf{Rank} & \multicolumn{2}{c|}{\textbf{Threat Category}}                                                                                         & \textbf{Base Score} & \textbf{Overall Score} & \textbf{Severity} \\ \hline
			1             & \multicolumn{1}{c|}{TC1} & \begin{tabular}[c]{@{}l@{}}Unauthorized SDN application \\ access with CSP user permissions\end{tabular}    & 9,0                 & 7,9                    & Critical          \\ \hline
			1             & \multicolumn{1}{c|}{TC2} & \begin{tabular}[c]{@{}l@{}}Unauthorized SDN controller access\end{tabular}                               & 9,0                 & 7,9                    & Critical          \\ \hline
			2             & \multicolumn{1}{c|}{TC3} & Man-in-the-middle                                                                                           & 8,9                 & 7,9                    & High              \\ \hline
			3             & \multicolumn{1}{c|}{TC4} & \begin{tabular}[c]{@{}l@{}}DoS - SDN controller in a \\ single controller setup\end{tabular}                & 6,8                 & 7,7                    & Medium            \\ \hline
			4             & \multicolumn{1}{c|}{TC5} & \begin{tabular}[c]{@{}l@{}}Unauthorized SDN application \\ access with tenant user permissions\end{tabular} & 6,5                 & 5,6                    & Medium            \\ \hline
			4             & \multicolumn{1}{c|}{TC6} & \begin{tabular}[c]{@{}l@{}}Unauthorized OpenFlow switch access\end{tabular}                              & 6,5                 & 4,6                    & Medium            \\ \hline
			5             & \multicolumn{1}{c|}{TC7} & \begin{tabular}[c]{@{}l@{}}Information disclosure of all \\ OpenFlow connections\end{tabular}               & 5,9                 & 6,7                    & Medium            \\ \hline
			5             & \multicolumn{1}{c|}{TC8} & \begin{tabular}[c]{@{}l@{}}Information disclosure of the \\ northbound interface\end{tabular}               & 5,9                 & 6,7                    & Medium            \\ \hline
			5             & \multicolumn{1}{c|}{TC9} & \begin{tabular}[c]{@{}l@{}}Information disclosure of the BGP \\ connection between controllers\end{tabular} & 5,9                 & 6,7                    & Medium            \\ \hline
			5             & \multicolumn{1}{c|}{TC10} & \begin{tabular}[c]{@{}l@{}}Information disclosure of data traffic\end{tabular}                           & 5,9                 & 6,7                    & Medium            \\ \hline
			6             & \multicolumn{1}{c|}{TC11} & DoS - OpenFlow switch                                                                                       & 4,0                 & 2,7                    & Medium            \\ \hline
			6             & \multicolumn{1}{c|}{TC12} & DoS - SDN application                                                                                       & 4,0                 & 3,5                    & Medium            \\ \hline
			7             & \multicolumn{1}{c|}{TC13} & \begin{tabular}[c]{@{}l@{}}Information disclosure of a \\ single OpenFlow connection\end{tabular}           & 3,7                 & 2,6                    & Low               \\ \hline
			7             & \multicolumn{1}{c|}{TC14} & \begin{tabular}[c]{@{}l@{}}DoS - SDN controller in a \\ multiple controller setup\end{tabular}              & 3,7                 & 2,6                    & Low               \\ \hline
		\end{tabular}%
	}
	\label{tab:cvss_scoring}
\end{table*}

\vspace{-15pt}
In the third step, threats with unpredictable severity, such as "Human Errors" from the PASTA analysis, are handled separately from the approach. These threats can be of any kind, making it challenging to define their impact. This unique treatment of the root threat does not imply its innocence. On the contrary, such characteristics make the threat even more dangerous. However, the risk and impact analysis cannot provide reliable results due to the unpredictability of the root threat. As a result, human errors are excluded from the severity analysis with the scoring tools, however, it is possible to minimize them by applying the presented measures in section 7 (Stage 4).

The approach to defining the TCs follows the reverse way compared to the PASTA analysis. During the PASTA assessments, root and sub-threats were determined, after which attack scenarios and vulnerabilities were derived from the threats. For categorizing security flaws, the effect on the SDN network is first defined, followed by threat correlation. This process yields 14 TCs mapped to three root threats. Subsequently, the CVSS methodology was used to rank the TCs based on their impact severity on a DC environment. As demonstrated in Table \ref{tab:cvss_scoring}, TCs with critical or high severity can potentially harm the entire DC environment and thus need to be mitigated first.

The results of the CVSS scoring, presented in Table \ref{tab:cvss_scoring}, support users of the SDN Security Evaluation Framework in classifying threats based on their impact severity on the SDN environment. This classification is necessary for prioritizing remediation and estimating the potential risk per TC. As previously described, TCs with critical or high severity can potentially harm the entire DC environment. This implies that DC operators must concentrate on remediation strategies to mitigate such threats. Suitable countermeasures to enhance SDN security are discussed in section 7 (Stage 4). TCs with medium or low severity have a lesser impact on the DC environment because they do not affect all SDN users/services, or they only compromise the confidentiality, integrity, or availability of data and services within the DC. Such threats are given lower priority for mitigation, or DC operators can choose to accept the risk posed by the threat. Nonetheless, it is recommended to implement countermeasures against all identified SDN threats to ensure the security of the entire DC environment. 
\vspace{-10pt}
\section{Attack Modelling (Stage 3)}
\label{sec:Stage3}
\vspace{-10pt}

In Stage 3, we modelled three attack scenarios for the three highest-ranked TCs from the previous stage. The Dictionary Attack, a well-known method for gaining unauthorized access to a target system (TC2), is an enhanced form of the Brute Force Attack. This attack aims to discover system user credentials by automatically trying every string combination. The time needed to find the correct credentials depends on password complexity and computing power. The Kali VM provides built-in tools, such as Patator, Medusa, THC Hydra, and Metasploit, to execute a brute force attack. Additionally, Kali provides a dictionary file named "rockyou," containing over 14 million common passwords. All mentioned tools can execute a dictionary attack using the "rockyou" file as input. Given that the Kali VM runs with 2024 MB of RAM and two CPUs, it's expected to take some time to find the correct password. Patator cracked the default password in 4 seconds (fastest), while THC Hydra took approximately 22 minutes (slowest).

The Man-in-the-middle Attack (TC3) poses a high-impact severity on a DC environment, according to Table \ref{tab:cvss_scoring}. This attack disrupts the assumption that users or devices communicate directly with the target system. Attackers position themselves between communication endpoints to intercept transmissions, potentially compromising credentials, stealing sensitive data, and providing different responses to the user or system. There are various kinds of man-in-the-middle attacks, such as session hijacking, replay attacks, IP spoofing, and eavesdropping attacks. For the experimental study, the focus was on the eavesdropping attack. This attack was executed on the Mininet VM, capturing packets for the ICMP, Telnet, and OpenFlow protocols. All traffic was analysed with Wireshark, and the SDN testbed, including network services, was exposed. The same attack also facilitated eavesdropping on an Telnet session, including login credentials.

\vspace{-20pt}
\label{sec:attack_scenarios}
\begin{figure*}[!h]
	\centering
	\includegraphics[width=\textwidth]{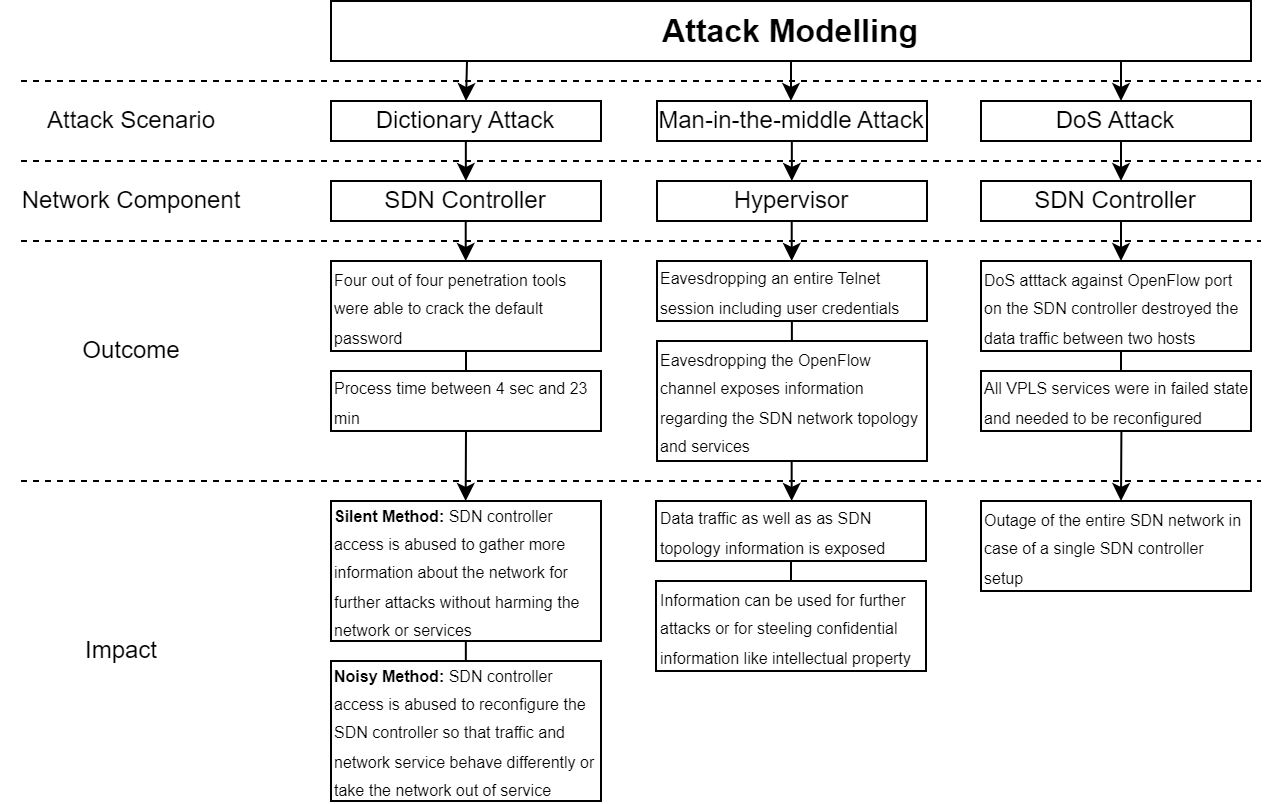}
	\caption{Executed Attack Scenarios on the SDN-based DC Environment Testbed.}
	\label{fig:attack_scenarios}
\end{figure*}
\vspace{-20pt}

The DoS attack against the SDN controller represents the last attack modelled in the experimental study. This attack represents TC4 from Table \ref{tab:cvss_scoring}, with medium severity in a single-controller setup. According to Carl et al. \cite{carl-2006}, there are two general types of DoS attacks. The first type, the vulnerability attack, uses malformed packets to exploit the weaknesses of the victim system or application. The second type, the flooding attack, involves the attacker sending the victim system a large and continuous network traffic workload, congesting legitimate workloads, and rendering services unavailable. A DoS attack in the form of an SYN flood attack was executed from the Kali VM against the OpenFlow port (6653) of the ONOS controller. With the help of the hping3 tool, over 4 million SYN packets were sent to the controller, interrupting traffic between all hosts within their VPLS after 8 seconds and terminating all three VPLS services. Due to the disrupted network services, the data traffic did not recover, requiring a VPLS service reconfiguration to restore the testbed to its initial state.

In summary, based on the risk and impact analysis presented in Table \ref{tab:cvss_scoring}, the three highest-ranked attacks (Dictionary Attack, Man-in-the-middle Attack, DoS Attack) are modelled for the experimental study. All the attacks executed in the experimental study against the SDN test setup are examples to illustrate the technical perspective of three attacks, explain their functionality, and verify their impact from the previous stage (Stage 2).

\vspace{-15pt}
\section{Threat \& Vulnerability Mitigation (Stage 4)}
\label{sec:Stage4}
\vspace{-10pt}

In Stage 4, the Security Evaluation Framework provides countermeasures for mitigating the threats and vulnerabilities identified and ranked in Stage 1. In this context, we have identified 18 mitigations (M1 – M18) that directly correspond to the threats (T1 – T18) identified in the first stage, as shown in Table \ref{mitigations}:

\vspace{-15pt}
\begin{table*}[!h]
	\caption{Mitigations of identified SDN threats (T1 - T18).}
	\centering
	\resizebox{\textwidth}{!}{%
		\begin{tabular}{|c|l|}
			\hline
			\textbf{Mitigations} & \multicolumn{1}{c|}{\textbf{Description}}                                                                                                                                                                                                                                                                                                                                                                                       \\ \hline
			M1                   & \begin{tabular}[c]{@{}l@{}}\tabitem only permit execution of signed scripts \\ \tabitem remove any unnecessary or unused shells or interpreters \\ \tabitem use application control where appropriate\end{tabular}                                                                                                                                                                  \\ \hline
			M2                   & \begin{tabular}[c]{@{}l@{}}\tabitem enable multi-factor authentication \\ \tabitem apply password policies according to Grassi et al. (2017) \\ \tabitem ensure proper privilege separation\end{tabular}                                                                                                                                                                           \\ \hline
			M3                   & \begin{tabular}[c]{@{}l@{}}\tabitem restrict file and directory permissions \\ \tabitem ensure proper user permissions are in place\end{tabular}                                                                                                                                                                                                                                                  \\ \hline
			M4                   & \begin{tabular}[c]{@{}l@{}}\tabitem enable multi-factor authentication \\ \tabitem apply password policies according to Grassi et al. (2017)\\ \tabitem restrict administrator accounts to as few individuals as possible\end{tabular}                                                                                                                                             \\ \hline
			M5                   & n/a*                                                                                                                                                                                                                                                                                                                                                                                                                             \\ \hline
			M6                   & \begin{tabular}[c]{@{}l@{}}\tabitem encrypt sensitive information, e.g. with SSL/TLS \\ \tabitem enable multi-factor authentication\end{tabular}                                                                                                                                                                                                                                                  \\ \hline
			M7                   & n/a*                                                                                                                                                                                                                                                                                                                                                                                                                             \\ \hline
			M8                   & \tabitem minimize the amount and sensitivity of data available to external parties                                                                                                                                                                                                                                                                                                                               \\ \hline
			M9                   & \begin{tabular}[c]{@{}l@{}}\tabitem implement access control mechanisms like ACLs \\ \tabitem log access control failures \\ \tabitem rate limit access to minimize the harm from automated attack tooling\end{tabular}                                                                                                                                                            \\ \hline
			M10                  & \begin{tabular}[c]{@{}l@{}}\tabitem identifiy sensitive data according to privacy laws, regulatory requirements, or business needs \\ \tabitem encrypt all data in transit with secure protocols such as TLS \\ \tabitem do not use legacy protocols such as File Transfer Protocol (FTP) and Simple Mail Transfer \\ Protocol (SMTP) for transporting sensitive data\end{tabular} \\ \hline
			M11                  & \begin{tabular}[c]{@{}l@{}}\tabitem use server-side input validation \\ \tabitem use SQL controls like LIMIT to prevent mass disclosure\end{tabular}                                                                                                                                                                                                                                              \\ \hline
			M12                  & \begin{tabular}[c]{@{}l@{}}\tabitem use threat modeling for critical authentication and access control \\ \tabitem limit resource consumption by user or service\end{tabular}                                                                                                                                                                                                                     \\ \hline
			M13                  & \begin{tabular}[c]{@{}l@{}}\tabitem establish a repeatable hardening process in an automated manner across multiple systems \\ \tabitem use ACLs to provide effective and secure separation between components or tenants\end{tabular}                                                                                                                                                           \\ \hline
			M14                  & \begin{tabular}[c]{@{}l@{}}\tabitem remove unused dependencies, unnecessary features, components, and files \\ \tabitem continuously inventory the versions of all components \\ \tabitem only obtain components from official sources over secure links\end{tabular}                                                                                                              \\ \hline
			M15                  & \begin{tabular}[c]{@{}l@{}}\tabitem enable multi-factor authentication \\ \tabitem do not use default credentials \\ \tabitem apply password policies according to Grassi et al. (2017) \\ \tabitem limit or increasingly delay failed login attempts and log all failures\end{tabular}                                                                             \\ \hline
			M16                  & \begin{tabular}[c]{@{}l@{}}\tabitem use digital signatures to verify the data is from the expected source \\ \tabitem ensure libraries are consuming trusted repositories \\ \tabitem ensure that there is a review process for configuration changes\end{tabular}                                                                                                                 \\ \hline
			M17                  & \begin{tabular}[c]{@{}l@{}}\tabitem ensure all login and access control failures can be logged with sufficient user context \\ \tabitem ensure log data is encoded correctly to prevent injections or attacks\end{tabular}                                                                                                                                                                        \\ \hline
			M18                  & \begin{tabular}[c]{@{}l@{}}\tabitem enforce “deny by default” firewall policies or network access control rules \\ \tabitem sanitize and validate all client-supplied input data \\ \tabitem do not send raw responses to clients\end{tabular}                                                                                                                                     \\ \hline
			\multicolumn{2}{l}{*This threat is difficult to mitigate with preventive controls since it is based on the exploit of system features}
		\end{tabular}%
	}
	\label{mitigations}
\end{table*} 

Table \ref{mitigations} offers specific countermeasures for each threat, based on the MITRE Corporation and OWASP. The suggested mitigations (M1 – M18) generally do not require a complex implementation of external tools or software. Some threats could be mitigated by simply activating the built-in- functionalities provided by the OS or protocols used in SDN. For example, the threat (T6) "Network Sniffing" can be mitigated by encrypting sensitive data (M6). Encryption via TLS is a built-in feature of the OpenFlow protocol that is not enabled by default, even though it is highly recommended to do so. Moreover, the man-in-the-middle attack from Stage 3 demonstrated how easy it is to eavesdrop on a not encrypted OpenFlow connection. 

For threats where corresponding mitigations are not applicable (M5 and M7), it's challenging to apply preventive controls because these threats exploit system features. However, that does not mean there are no countermeasures available at all. Mitigations in the second category can protect the SDN from such threats using a central solution. Rather than applying countermeasures against each identified SDN threat, a more generic solution is implemented to safeguard the network against several types of attacks on multiple SDN components. Typically, these solutions entail a complex integration of external systems.

\vspace{-15pt}  
\subsection{Policy-based SDN Security Architecture}

One of the solutions to mitigate several types of attacks is the PbSA a security application implemented in the northbound interface of the SDN controller, called PbSA. The following figure shows the presented PbSA \cite{varadharajan-2019}:

\vspace{-15pt} \label{sec:pbsa} \begin{figure*}[!h] \centering \includegraphics[width=\textwidth]{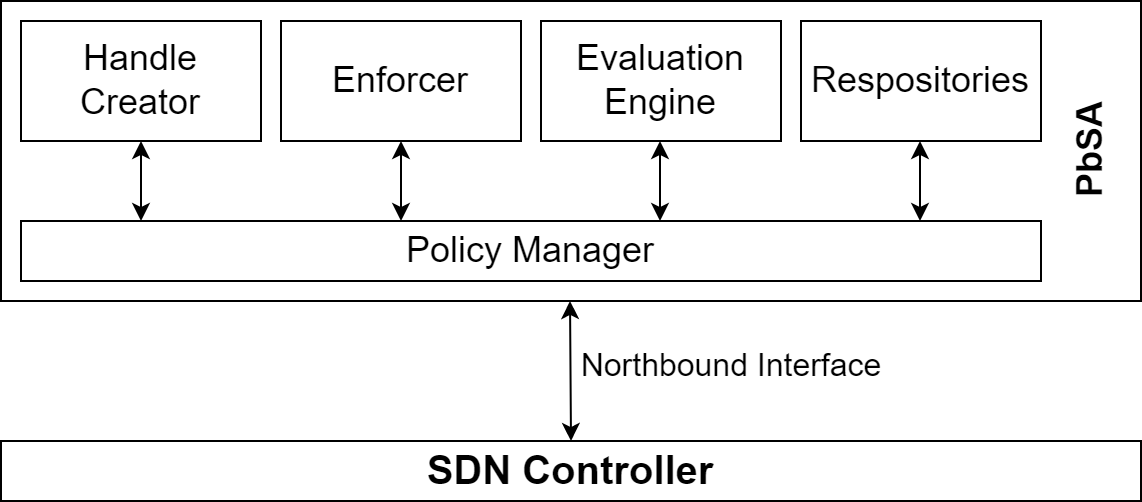} \caption{Policy-based Security for SDN (adapted from Varadharajan et al., 2019)} \label{fig:pbsa} \end{figure*}
\vspace{-15pt}  

The primary objective of the PbSA is to authorize flows in an SDN environment based on security policies. These security policies utilize different flow parameters to authorize flows within the SDN. The policy administrator can specify finely grained security policies based on various attributes such as users, devices, or Autonomous Systems (AS). Furthermore, PbSA enforces a default deny policy, which drops flow requests that are not explicitly permitted. Moreover, the solution contains an additional module to secure the communication between network devices with the help of distributed keys. A novel feature includes path-based policies, which establish an alternative path through the SDN when the original path is blocked (during a DoS attack). 

PbSA mitigates several threats across multiple SDN layers. For instance, it can detect DoS attacks, block suspicious flows, and redirect data traffic over alternative paths. Additionally, the architecture can block all traffic that does not satisfy a permit policy by default (e.g., dropping malicious flows or denying misconfigured services), representing an additional security layer in the form of an internal SDN firewall. Moreover, the new module ensures secure communication between several network devices by mitigating the "Network Sniffing" threat (T6), enhancing the SDN security for the control and data plane, and the southbound interface.

\subsection{Secure SDN based on Blockchain}

Another centralized solution was proposed by Weng et al. \cite{weng-2019}, which is based on a Blockchain layer between the SDN controller and the data plane. Attribute-based encryption is used to ensure fine-grained access control for encrypted data at the northbound interface. For communication between the Blockchain layer and the data layer, the HOMQV protocol is established. The Blockchain layer provides resource-sharing and resource-recording functionalities among multiple SDN controllers on the control plane. Hence, the Blockchain is used to record the network resources of each controller and to share recorded network events among all controllers, ensuring a uniform network view. The following figure shows the SDN topology, extended by a Blockchain layer \cite{weng-2019}:

\vspace{-15pt} \label{sec:blockchain} \begin{figure*}[!h] \centering \includegraphics[width=4cm]{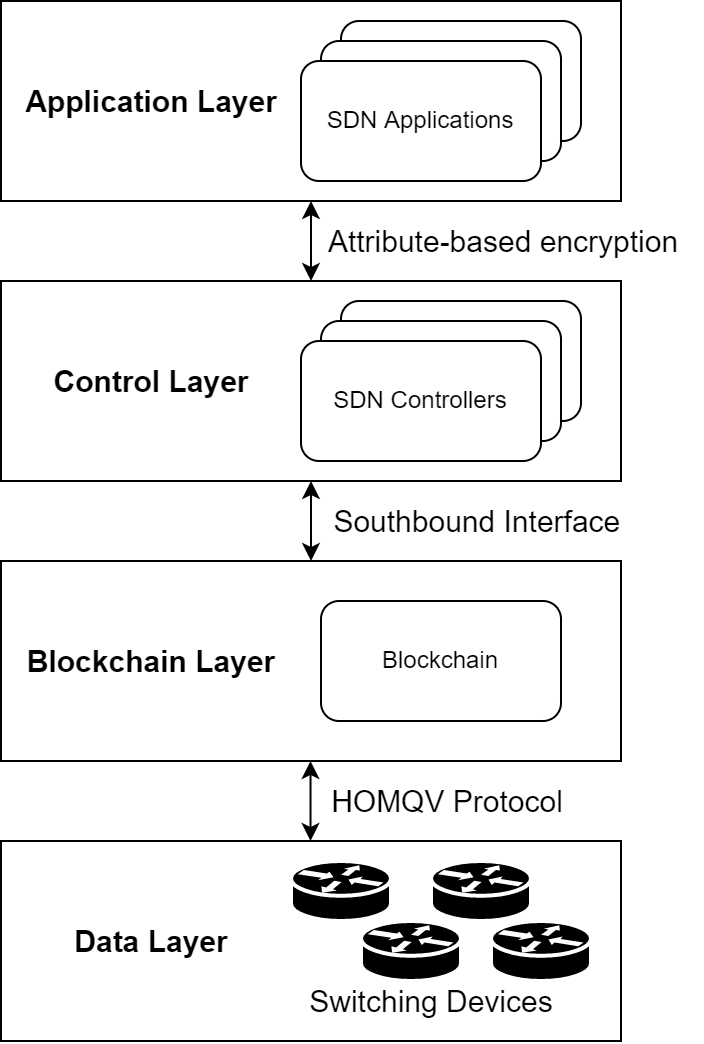} \caption{SDN Topology with Blockchain Layer (adapted from Weng et al., 2019)} \label{fig:blockchain} \end{figure*}

\vspace{-15pt}  

The Blockchain records all application flows and network events associated with the respective network conditions. Such information is stored as raw transactions in the Blockchain. Furthermore, smart contracts are used to implement security protocols, for instance, to alert of an SDN controller failure. In addition, the controllers participating in the underlying Blockchain send their recorded network data (from the application and data layers) as raw transactions into the Blockchain. This makes the Blockchain layer a real-time reliable instance of all recorded application flows and all time-series of the network-wide views. The proposed solution provides a unified security mechanism for each SDN component. It can also mitigate threat T5, which cannot be easily mitigated with preventive controls since it is based on the abuse of system features. However, due to the introduced Blockchain layer and HOMQV protocol, this security mechanism provides secure authentication for applications, controllers, and switches.

\vspace{-15pt}
\subsection{A Distributed SDN Framework for Scalable Network Security}

The overall system architecture of another central solution for mitigating several threats, called TENNISON, is shown in the following figure:

\vspace{-15pt} \label{sec:tennison} \begin{figure*}[!h] \centering \includegraphics[width=11cm]{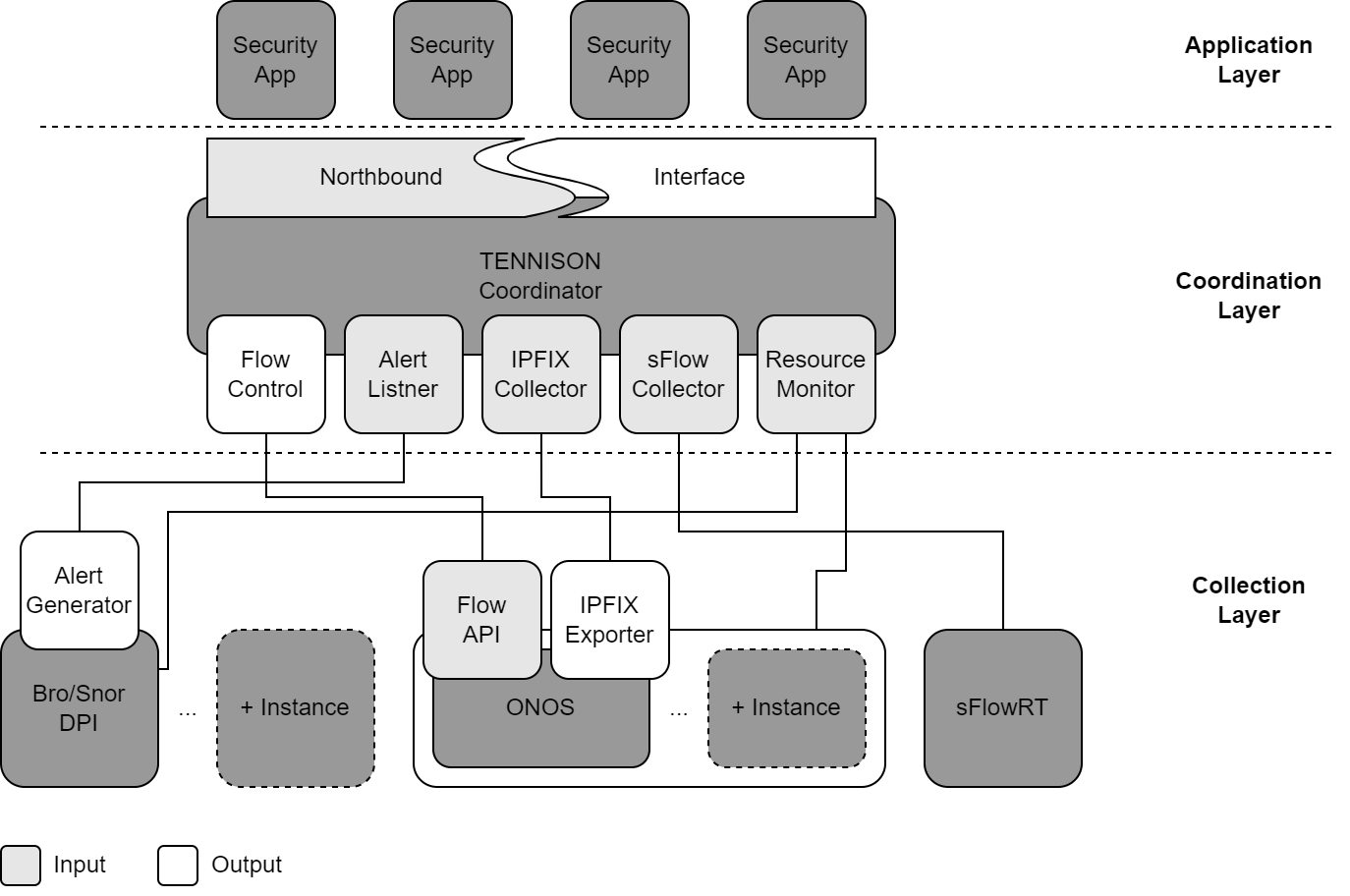} \caption{TENNISON System Architecture (adapted from Fawcett et al., 2018).} \label{fig:tennison} \end{figure*}
\vspace{-15pt}

TENNISON was developed by Fawcett et al. \cite{fawcett-2018}, who provided a multi-level, distributed monitoring and remediation SDN framework for scalable network security. The TENNISON framework presents an adaptive and extensible security platform that is technology-independent and capable of supporting a wide range of security functions. Mainly focusing on enhancing the security of the SDN control and data plane, the TENNISON security framework is a comprehensive solution. The distributed ONOS controllers ensure availability and resiliency, and the monitoring and remediation functionality detects attacks and drops malicious traffic in the data plane. Fawcett et al. \cite{fawcett-2018} demonstrated the effectiveness of their solution for four different attack types: DoS, DDoS, scanning, and intrusion. This security framework can also manage the challenges of mitigating threat T7, which cannot be easily mitigated using specific countermeasures.

\subsection{Summary}

In the final stage of the Security Evaluation Framework (Stage 4), we demonstrated how to mitigate the identified SDN threats by providing two different methodologies. The first method involves mitigating specific SDN threats by utilizing built-in OS or protocol features to enhance resiliency against such threats. Table \ref{mitigations} enumerates the countermeasures against nearly all identified SDN threats as determined in section \ref{sec:Stage1}. This process generally does not require the implementation of external tools or software that mightintroduce new SDN vulnerabilities.

The second method involves implementing a more centralized security solution to bolster the robustness of the SDN network against multiple threats across various components. Such a solution often requires the implementation of external systems, which may introduce new vulnerabilities to the network. We presented three different central solutions for enhancing SDN security.

\begin{figure*}[!h] \centering \includegraphics[width=\textwidth]{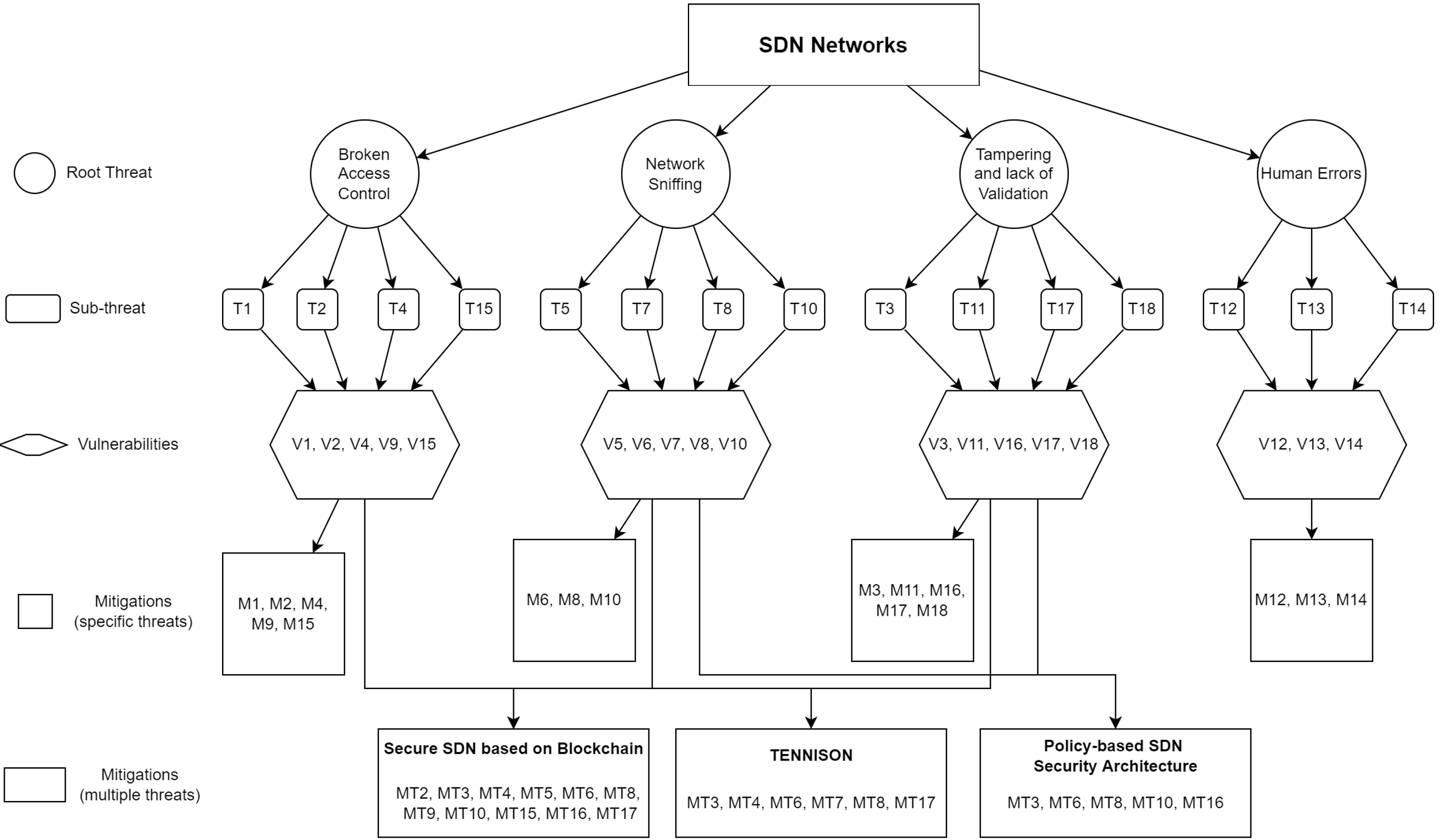} \caption{"Threat-Vulnerability-Mitigation"-Correlation Map (Ivki'c et al., 2023).} \label{fig:correlation-map} \end{figure*}

To comprehend the provided countermeasures in conjunction with the threat analysis, a well-structured map would provide a correlation between SDN threats, vulnerabilities, and mitigations. As shown in Figure \ref{fig:correlation-map}, a threat tree is used to show (in a top-down manner) the root threat with sub-threats, the associated vulnerabilities, and mitigations. The root threat is depicted in a circular shape, and all descendant nodes represent sub-threats. If any node has multiple child nodes, their relation is disjunctive by default unless otherwise marked. This article includes an additional hierarchy level to link the vulnerabilities to the associated threats. Moreover, each vulnerability is again linked with the associated mitigations (M1 - M18) and/or mitigating central solutions. Figure \ref{fig:correlation-map} displays the final output (Output 4) of the Security Evaluation Framework:

\vspace{-15pt}
\label{sec:correlation-map} 

\vspace{-35pt}
\section{Conclusions and Future Work}
\label{sec:conclusion}

The operation of cloud applications for a wide range of users is enhanced using SDN in DC environments over traditional networking techniques. By separating the control plane from the data plane, the more advanced SDN architecture provides a network solution that is flexible, manageable, and adaptable. However, it also introduces new network security threats and risks.

This article presents an SDN Security Evaluation Framework and demonstrates how it can be used to identify SDN-related threats and vulnerabilities. Upon detecting security threats, the framework suggests conducting a risk and impact analysis including rating the potential impact on DC environments if no countermeasure is taken. Followed by that attack scenarios are modelled next, to verify the calculated impact severity and to find mitigating measures to increase the overall SDN security. The resulting correlation mapping of the framework provides an overview of all threats, vulnerabilities, and mitigating countermeasures. The framework provides all the necessary models, tools, and methodologies to enable the extension of the correlation mapping and impact classification even for vulnerabilities that have not been identified yet. Both the presented framework and the results from this paper can assist network administrators to evaluate, categorise, compare, and improve the security of their SDN-based DCs.

For future work plan to extend the SDN Security Evaluation Framework by providing a comprehensive list of attack scenarios. Given the large number of possible attacks per threat, a complete list would help to better understand their technical implementation and to develop solutions to mitigate them.

\end{document}